\documentclass{article}

\usepackage{arxiv}

\usepackage[utf8]{inputenc} 
\usepackage[T1]{fontenc}    
\usepackage{hyperref}       
\usepackage{url}            
\usepackage{booktabs}       
\usepackage{amsfonts}       
\usepackage{nicefrac}       
\usepackage{microtype}      
\usepackage{lipsum}		
\usepackage{graphicx}
\usepackage{natbib}

\title{Holding-based evaluation upon actively managed stock mutual funds in China}

\author{ \href{https://orcid.org/0000-0001-7431-1619}{\includegraphics[scale=0.06]{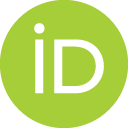}\hspace{1mm}Huimin Peng}\thanks{This piece of work is done during my Post-doctor research at PBCSF, Tsinghua University. Great thanks to Professor Quanwei Cao for helpful comments. Great thanks to my advisor Professor Zhuo Chen for advising me on this research project.} \\
	\texttt{hpeng2@ncsu.edu} \\
\texttt{penghuimin@inspur.com} \\
	\texttt{peng.huimin.pennie@gmail.com} \\
}

\begin{document}
\maketitle

\begin{abstract}
We analyze actively managed mutual funds in China from 2005 to 2017. We develop performance measures for asset allocation and selection. We find that stock selection ability from holding-based model is positively correlated with selection ability estimated from Fama-French three-factor model, which is price-based regression model. We also find that industry allocation from holding-based model is positively correlated with timing ability estimated from price-based Treynor-Mazuy model most of the time. We conclude that most actively managed funds have positive stock selection ability but not asset allocation ability, which is due to the difficulty in predicting policy changes.
\end{abstract}

\keywords{Holding-based model \and Portfolio management \and Brinson model}

\section{Introduction}

This paper applies Brinson model proposed in \citet{Brinson1995} to study the performance attribution of actively managed stock mutual funds in China. It uses fund holdings from 2005 to 2017 to calculate performance measures, such as asset allocation, industry allocation and within-industry stock selection.
Our objective is to evaluate active management through the decomposition of excess return using Brinson model.

We make three contributions to literature. First, we provide Brinson model results of stock funds in China from 2005 to 2017. We compare fund holdings with fund benchmark to evaluate the constituents of active management.
Second, we analyze whether performance measures are persistent for each fund over time. We conclude that stock selection is better in persistence than industry allocation. 
Third, we show that stock selection measure from Brinson model is positively associated with selection from Fama-French three-factor model. Brinson model is based upon fund holdings and stock return data, but Fama-French three-factor model is dependent upon the time series of fund's adjusted net asset value. Though Brinson model and Fama-French three-factor model are based upon different data, the corresponding performance measures are positively correlated.

We calculate performance measures of stock mutual funds using Brinson model \citep{Brinson1995}. We analyze actual stock and bond holding proportions, compare them with stock and bond weights reported in benchmark and see whether it holds more stocks when market is bullish and less stocks when market is bearish.
We study the industries held by funds and industries in benchmark to check whether it holds more in bullish industries and less in bearish ones. 
We also calculate whether funds hold more in stocks with better performance from each industry and less in stocks with poor performance.

Our conclusion is threefold. First of all, most funds have positive within-industry stock selection, but not positive asset allocation. On average, within-industry stock selection makes the largest contribution to excess return, and industry allocation is the second largest component. 
Second, within-industry stock selection is the best in persistence among the performance measures we calculate. 
Stock funds do not show persistence in asset allocation.
Last but not the least, within-industry stock selection is positively correlated with Alpha (selection ability) from Fama-French three-factor model. And industry allocation is positively correlated with timing ability from Treynor-Mazuy model most of the time.

This is the outline of the paper.
Section \ref{sec:Literature} positions our research in the stock fund literature concerning performance attribution. 
Section \ref{sec:Model} introduces Brinson model and performance measures of stock funds. 
Section \ref{sec:Data} provides data summary. 
Section \ref{sec:Persistence} analyzes the persistence of performance measures from Brinson model. 
Section \ref{sec:Association} presents the association study between selection from Brinson model and Fama-French three-factor model, and between timing from Brinson model and Treynor-Mazuy model. 
Section \ref{sec:Conclusion} outlines the conclusions.

\section{Related Literature} \label{sec:Literature}

There are two classes of models for analyzing the performance attribution of mutual funds: one is based upon decomposition of excess return; the other applies time series regression.

Our paper considers performance attribution of actively managed stock fund in China from the angle of performance decomposition.
We apply Brinson model \citep{Brinson1995} to evaluate active management ability of Chinese stock funds.
Brinson model decomposes excess return into within-sector stock selection, sector allocation and interaction term \citep{Brinson1995}. Details on Brinson model are further discussed in \citet{Singer1991}. 

We are among the very few literature which apply Brinson model to analyze Chinese stock mutual fund performance.
We consider fund holdings and benchmark for the attribution of fund's excess return. 
Although \citet{Wu2014} argue that active management struggles to outperform market which explains the success of index investing in recent years, we show that in the long run, active management brings excess return for funds. In \citet{Arbaa2017}, it describes that timing ability in US equity fund is not prominent, neither in Israel. We find that industry allocation from Brinson model is positive for more than half stock funds in China. 

There are many related literature on the topic of performance evaluation using holdings data.
For example,
\citet{Griffin2009How} use hedge fund holdings to analyze the return of hedge fund, and find that hedge fund does not show superior within-sector stock selection or superior sector allocation.
In our findings, most stock mutual funds demonstrate superior within-industry stock selection.
\citet{Gelos2005} apply fund holdings to construct a measure on country transparency and conclude that funds show a tendency to invest more in more transparent countries and less in less transparent ones.
In addition, researchers raise the issue that unobserved behavior of funds help explain the source of excess return as well, as in \citet{Kacperczyk2008}. For now, we only consider fund holdings data which is observable. 

On the other hand, we study the performance persistence of within-industry stock selection, industry allocation and asset allocation using techniques in \citet{CarhartMarkM1997}.
We demonstrate that the performance persistence of within-industry stock selection is better than that of industry allocation. 
\citet{Bollen2005} conclude that persistence is a short-term phenomenon for funds within the time window of less than one year. We analyze performance persistence between every semi-annual time window as well.
In \citet{Fu2017}, it analyzes the performance persistence of selection and timing for winner funds specifically. \citet{Boyson2008} present that young hedge funds with small size are most significant in performance persistence. 

We know that researchers also develop regression models to account for performance attribution of actively managed fund, such as \citet{Fama1992} and \citet{Mazuy1966}.
Although \citet{Chan2009} point out that fund performance attribution results are very different across various evaluation methods, we find that the association between selection from Brinson model and selection from Fama-French three-factor model are positively correlated. We also find positive association between industry allocation from Brinson model and timing from Treynor-Mazuy model. 

Another performance measure we should consider is
asset allocation, since its contribution may be more important than active management in respect of driving excess return, as pointed out in \citet{Xiong2010}. As a result, we consider asset allocation using a definition similar to industry allocation from Brinson model to see its performance.
\citet{Kowara2013} demonstrate that asset allocation using risk factors is inherently comparable in performance to asset allocation using asset classes. 
In our paper, we define asset allocation in terms of asset classes.

Last but not the least,
fund holdings are powerful tools for making predictive analysis on funds and stocks future performance. For example, 
\citet{Wermers2012} use fund holdings to develop a performance measure which can predict stock return in US. In \citet{Elton2010}, it points out that 
selection derived from fund holdings data is more powerful in predicting future performance than selection obtained using time series regression. It also demonstrates that prediction works better when we increase the time frequency of fund holdings data. \citet{Cremers2009} develop a performance measure named active share from fund holdings which can be applied to predict future fund performance. Funds whose active share is highest show not only superior performance in the future but also better performance persistence. In light of these phenomena found in previous literature, we also study the out-of-sample properties of these performance measures and find that within-industry stock selection is the best in predicting future fund returns.

\section{Model} \label{sec:Model}

In this section, we present the performance measures from Brinson model \citep{Brinson1995}, which decomposes excess return into industry allocation, within-industry stock selection and interaction term. We calculate asset allocation using definitions similar to industry allocation. 
These performance measures are used in analysis from section \ref{sec:Persistence} to \ref{sec:Association}.

\subsection{Performance Measures from Brinson Model}
\label{sec:Performance_measures}

\begin{figure}[!htb]
	\centerline{\includegraphics[width=7in]{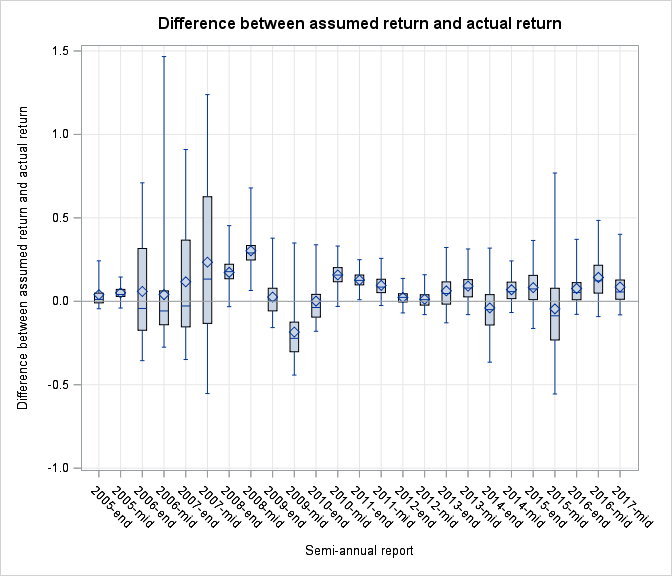}}
	\noindent\caption{}
	{\bf Box-plot of fund's assumed return and actual return in Brinson model for six months before report.} 
	
	In Brinson model, we assume that fund's holding remains the same as reported for half a year.
	Here we study two assumptions: fund's holding is valid for six months after the report, and fund's holding is valid for six months before the report.
	Under the assumption that fund's holding is valid for six months before the report, we calculate assumed return for funds. We also observe fund's actual return in six months before the report. We calculate the difference between assumed return and actual return. Then we plot box-plots for all funds over all semi-annual reports. In box-plot, the box shows the interval between 25\% percentile and 75\% percentile. The outside bars show the interval between 2.5\% percentile and 97.5\% percentile. \label{fig:diffasp06}
\end{figure}

\citet{Brinson1995} propose a decomposition model which is based upon fund holdings and benchmark constituents.
Consider $w_{ib}$ to be industry $i$ weight in benchmark, 
$r_{ib}$ as overall return from all industry $i$ stocks in benchmark, $w_{if}$ to be industry $i$ proportion in fund holdings, and $r_{if}$ as overall return from all industry $i$ stocks held by funds. Consider the number of industries covered by fund holdings and benchmark constituents to be $n$.
Returns $r_{ib}$ and $r_{if}$ correspond to six-month time window covering each semi-annual report. 
For example, for 2017 mid-year report, we may examine 2017 Jan to 2017 June which implies six months before the report. Or we may study the report using data from 2017 June to 2017 December which implies six months after the report.
Or we can examine any six-month period covering 2017 mid-year report. 
Under Brinson model, we assume that 2017 mid-year stock holding is valid in the chosen six-month period covering 2017 June.
In figure \ref{fig:diffasp06}, we examine this assumption empirically. We calculate the actual return of funds in six months before the report, and the assumed return under the reported stock holding. Then we take the difference between actual return and assumed return. For all the funds, we plot box-plot. In the box-plot, end bars are 2.5\% and 97.5\% percentiles which form 95\% confidence interval. 
The null hypothesis is that there is no difference between actual return and assumed return.
For most semi-annual reports, the 95\% confidence interval covers zero so that we cannot reject the null hypothesis. 

\bigskip
\centerline{\bf [Place Figure~\ref{fig:diffasp06} about here]}
\bigskip

\begin{figure}[!htb]
	\centerline{\includegraphics[width=7in]{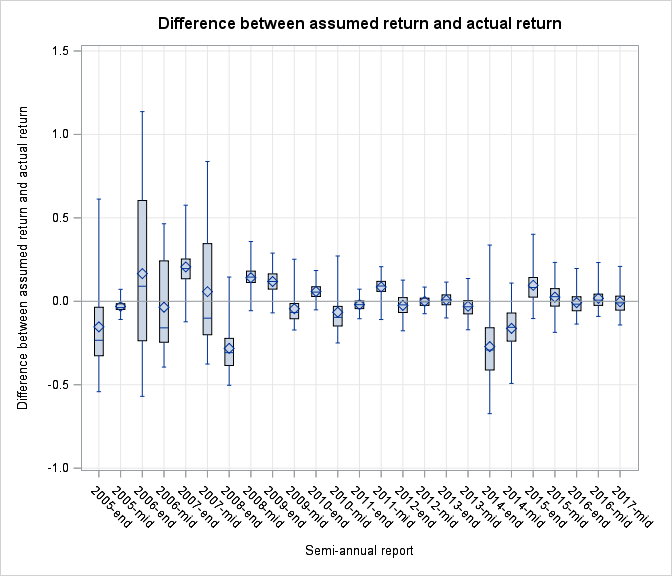}}
	\noindent\caption{}
	{\bf Box-plot of fund's assumed return and actual return in Brinson model for six months after report.} 
	
	In Brinson model, we assume that fund's holding remains the same as reported for half a year.
	Here we study two assumptions: fund's holding is valid for six months after the report, and fund's holding is valid for six months before the report.
	Under the assumption that fund's holding is valid for six months after the report, we calculate assumed return for funds. We also observe fund's actual return in six months after the report. We calculate the difference between assumed return and actual return. Then we plot box-plots for all funds over all semi-annual reports. In box-plot, the box shows the interval between 25\% percentile and 75\% percentile. The outside bars show the interval between 2.5\% percentile and 97.5\% percentile. \label{fig:diffasp612}
\end{figure}

Similarly, in figure \ref{fig:diffasp612}, we study the box-plot for the difference between actual return and assumed return. Here the assumption is that reported stock holding is valid for six months after the report rather than before the report as in figure \ref{fig:diffasp06}. 
We can see that in figure \ref{fig:diffasp612}, all 95\% confidence intervals cover zero. In figure \ref{fig:diffasp06}, only 2008 mid year report and 2011 end year report do not cover zero. Most 95\% confidence intervals cover zero.

As a result, from our empirical analysis, no matter which six-month period we choose, Brinson model assumption holds approximately. We can apply performance measures from Brinson model to analyze the performance attribution of stock mutual funds in China.

\bigskip
\centerline{\bf [Place Figure~\ref{fig:diffasp612} about here]}
\bigskip

We define the following performance measures based upon Brinson model.
First, within-industry stock selection measures whether funds hold more in stocks with superior performance in all industries and less in inferior ones.
Within-industry stock selection is defined as 
\begin{equation} \label{eq:wiss}
SS =  \sum_{i=1}^{n} \left(r_{if}-r_{ib}\right)w_{ib}.
\end{equation}
Second, industry allocation shows whether funds hold more in industries with better performance and less in worse ones.
Industry allocation is defined as
\begin{equation} \label{eq:ia}
IA =  \sum_{i=1}^{n} r_{ib}\left(w_{if}-w_{ib}\right).
\end{equation}
Finally, interaction term describes whether funds simultaneously hold more in better industries and better within-industry stocks.
Interaction term is defined as
\begin{equation} \label{eq:it}
IT =  \sum_{i=1}^{n} \left(r_{if}-r_{ib}\right)\left(w_{if}-w_{ib}\right).
\end{equation}
Under the assumption that benchmark reflects the investment protocol of funds, we conclude that industry allocation and within-industry stock selection are mutually independent conditional upon benchmark performance. 

Consider $w_{fs}$ to be stock proportion in fund holdings, $w_{bs}$
to be stock weight in benchmark, and $r_{s}$ to be stock market return. Consider $w_{fb}$ as bond proportion in fund holdings, $w_{bb}$ as bond weight in benchmark, and $r_{b}$ as bond market return. Returns $r_s$ and $r_b$ correspond to three-month time window after each quarterly report. 
We study the asset allocation assuming that asset allocation holds for three months after each quarterly report. 
Asset allocation is defined as
\begin{equation} \label{eq:aa}
AA = r_{s}\left(w_{fs}-w_{bs}\right)
+r_{b}\left(w_{fb}-w_{bb}\right).
\end{equation}
Asset allocation is defined in a similar way to industry allocation. Asset allocation describes whether funds invest more in stocks when market is bullish and less in stocks when market is bearish.

\subsection{Regression Model}
\label{sec:Regression_Model}

In our analysis, we apply performance measures from Brinson model to evaluate the stock funds. Another commonly used way to assess fund performance is to utilize regression model to depict the selection and timing ability of these funds. 
For example, \citet{Fama1992} propose Fama-French three-factor model as follows
\begin{equation} \label{eq:fama_french3}
r_{t}-r_{f}=\alpha+\beta_m (r_{mt}-r_{f})+\beta_{smb}SMB_t+\beta_{hml}HML_t+\varepsilon_{t},
\end{equation}
where $r_f$ is risk-free rate and $\alpha$ is selection.
For each fund, we apply model (\ref{eq:fama_french3}) to estimate the intercept $\hat{\alpha}$, which is selection ability of this fund.
On the other hand,
\citet{Mazuy1966} propose Treynor-Mazuy model as follows
\begin{equation} \label{eq:Treynor_Mazuy}
r_{t}-r_{f}=\alpha+\beta_m (r_{mt}-r_{f})+\gamma (r_{mt}-r_{f})^2+\varepsilon_{t},
\end{equation}
where $\gamma$ is timing. Similarly we apply model (\ref{eq:Treynor_Mazuy}) to each fund to estimate its timing ability.
Selection from model (\ref{eq:fama_french3}) describes fund's ability to invest in superior stocks. Timing from model (\ref{eq:Treynor_Mazuy}) depicts fund's ability to invest more in stocks when market is bullish and less when market is bearish. 
In our performance measures derived from Brinson model, we can see that within-industry selection implies similar ability of funds to selection from model (\ref{eq:fama_french3}), and that industry allocation implies similar ability to timing from model (\ref{eq:Treynor_Mazuy}). 
Although regression model is based upon time series and Brinson model is based upon fund's holding data, these corresponding performance measures should be positively correlated. In section \ref{sec:Association}, we analyze the association between these related performance measures from different performance evaluation models.

\subsection{Benchmark Model}
\label{sec:Benchmark_Model}

Benchmark reflects the investment protocol of stock funds. We do not observe all benchmark to be good fit for measuring fund's performance. In some period when benchmark significantly under-performs the market, fund's actual holdings may deviate to be completely different from benchmark constituents. As a result, we use the following regression model to test whether benchmark is proper:
\begin{equation} \label{eq:benchmark_check}
r_{t}=\alpha+\beta_d r_{dt}+\varepsilon_{t},
\end{equation}
where $r_t$ is return of fund and $r_{dt}$ is return of the fund's benchmark.
To test whether benchmark is appropriate, we use the following hypothesis:
\begin{equation} \label{eq:benchmark_check_hypothesis}
H_0:\beta_d=1~vs~H_a:\beta_d\neq 1.
\end{equation}
We argue that benchmark is not appropriate for the fund if we reject the null hypothesis.

On the other hand, we also compute the average and standard deviation of the time series $(r_t-r_{dt})$, $t=1,2,\cdots,T$ so that we can see whether fund return tracks benchmark return well.
A benchmark regression model widely used in literature is as follows
\begin{equation} \label{eq:benchmark_reg1}
r_{t}=\alpha+\beta_d r_{dt}+\beta_{smb}SMB_t+\beta_{hml}HML_t+\varepsilon_{t}.
\end{equation}
Under model in (\ref{eq:benchmark_reg1}), it is assumed that benchmark, size and value are
independent characteristics of fund. 
Under model (\ref{eq:benchmark_reg1}), we also test 
hypothesis in (\ref{eq:benchmark_check_hypothesis}) to see whether benchmark is appropriate or not.

Selection $\alpha$ from model (\ref{eq:benchmark_reg1})
is different from selection $\alpha$ from Fama-French three-factor model. In section \ref{sec:Association}, we will see that selection $\alpha$ from model (\ref{eq:benchmark_reg1}) is positively correlated with selection $\alpha$ from Fama-French three-factor model. Selection $\alpha$ from Fama-French three-factor model is positively correlated with within-industry stock selection from Brinson model.
Another widely applied benchmark regression model is posited to be
\begin{equation} \label{eq:benchmark_reg2}
r_{t}-r_{dt}=\alpha+\beta_m r_{mt}+\beta_{smb}SMB_t+\beta_{hml}HML_t+\varepsilon_{t},
\end{equation}
where $r_{mt}$ is overall market return.
Under model in (\ref{eq:benchmark_reg2}),
it is assumed that benchmark is well designed for the fund.
We analyze the performance attribution of the excess return when fund is compared to its benchmark rather than the risk-free rate. From our empirical analysis, selection from model 
(\ref{eq:benchmark_reg2}) is almost the same as selection from Fama-French three-factor model.

\subsection{Performance Persistence Model}
\label{sec:Performance_Persistence_Model}

After we derive performance measures, we analyze whether these performance measures are persistent over time for funds. 
We use the following model to test whether performance measures are persistent or not.
\begin{equation} \label{eq:persistence_check}
r_{t}=\alpha+\beta_1 r_{t-1}+\varepsilon_{t},
\end{equation}
where $r_t$ is return of performance measure in period $t$ and $r_{t-1}$ is return of performance measure in period $t-1$. We formulate the hypothesis as follows.
\begin{equation} \label{eq:persistence_check_hypothesis}
H_0:\beta_1=0~vs~H_a:\beta_1>0.
\end{equation}
We agree that the fund's performance measure shows significantly positive persistence if we reject the null hypothesis, which implies that return of performance measure is positively correlated with previous return. 
In section \ref{sec:Persistence}, we provide a summary on performance persistence of these performance measures. 

\section{Data} \label{sec:Data}

We present the database for calculating performance measures and for conducting the following analysis.
Open-end actively managed stock mutual fund in China dates back to 2005 when only about 50 such funds are in the market. 
Before 2005, there are not many open-end stock fund. For close-end funds, price discount and lack of liquidity can both be issues which should be considered \citep{Cherkes2009}. In our paper, we consider only open-end stock mutual fund under active management from 2005 to 2017.

\subsection{Data Summary} \label{sec:Data_Summary}

\begin{figure}[!htb]
	\centerline{\includegraphics[width=7in]{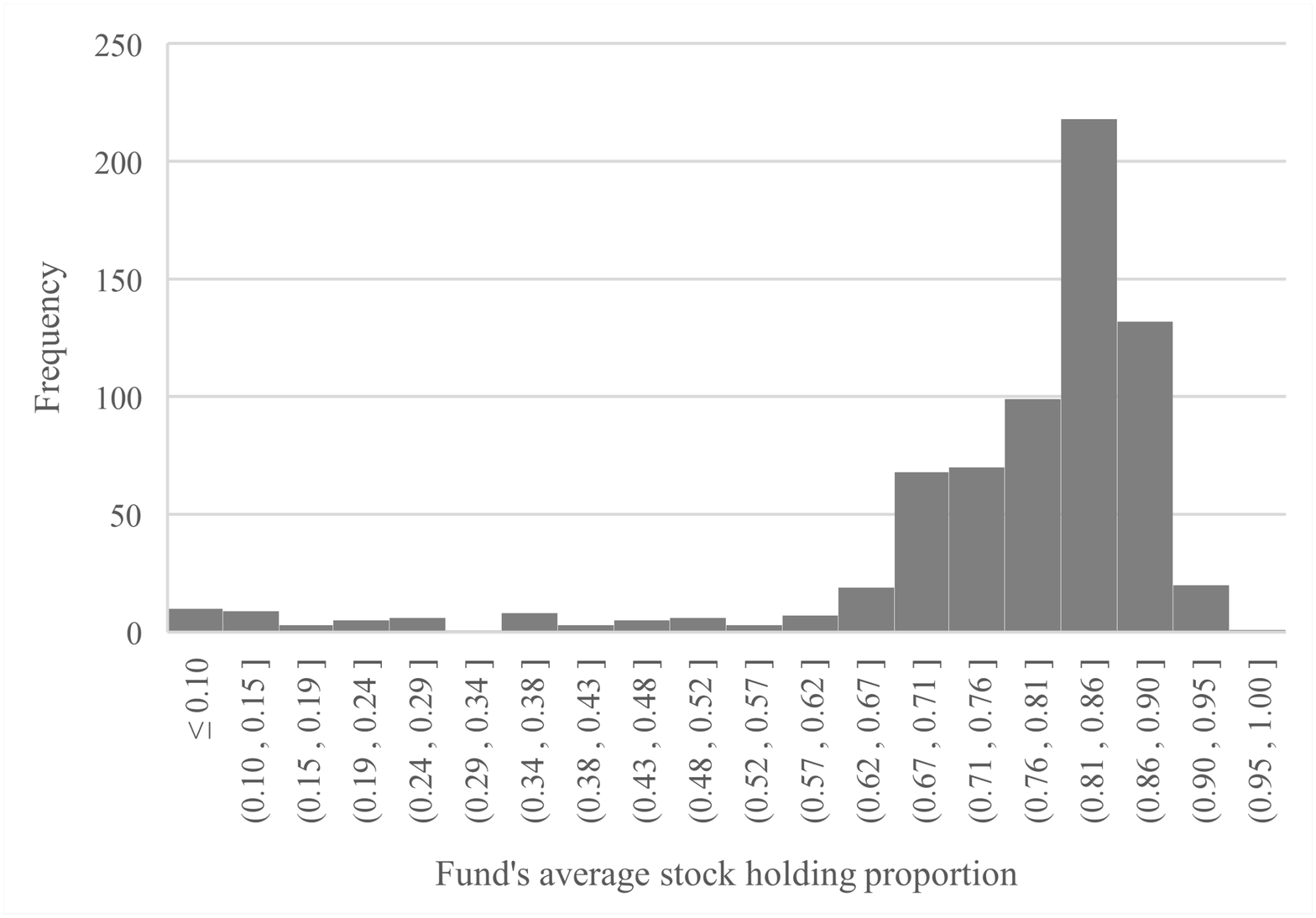}}
	\noindent\caption{}
	{\bf Histogram of fund's average stock holding proportion.} 
	
	This figure describes the proportion of stock holding within all fund holdings. We consider stock funds with continuous performance data from 2015 to 2017. In this three-year sample, 692 stock funds are present.
	For each fund, we calculate its average stock holding proportion from 2015-2017 reports. Then we plot histogram over all funds.  \label{fig:1}
\end{figure}

In China, fund asset allocation data are available at quarterly frequency. We can see the proportion of stocks and bonds held by the fund at each quarter. We download quarterly fund asset allocation data from Wind. 
Figure \ref{fig:1} provides a histogram of fund's average stock holding proportion. We examine funds with continuous performance from 2015 to 2017. In this three-year sample, we calculate average stock holding proportion of each fund. Most stock funds hold more than 60\% of stocks in their holdings. More than half stock funds hold more than 80\% of stocks in their holdings.
Stock funds contain a great proportion of stocks in their holdings.

\bigskip
\centerline{\bf [Place Figure~\ref{fig:1} about here]}
\bigskip

\begin{figure}[!htb]
	\centerline{\includegraphics[width=7in]{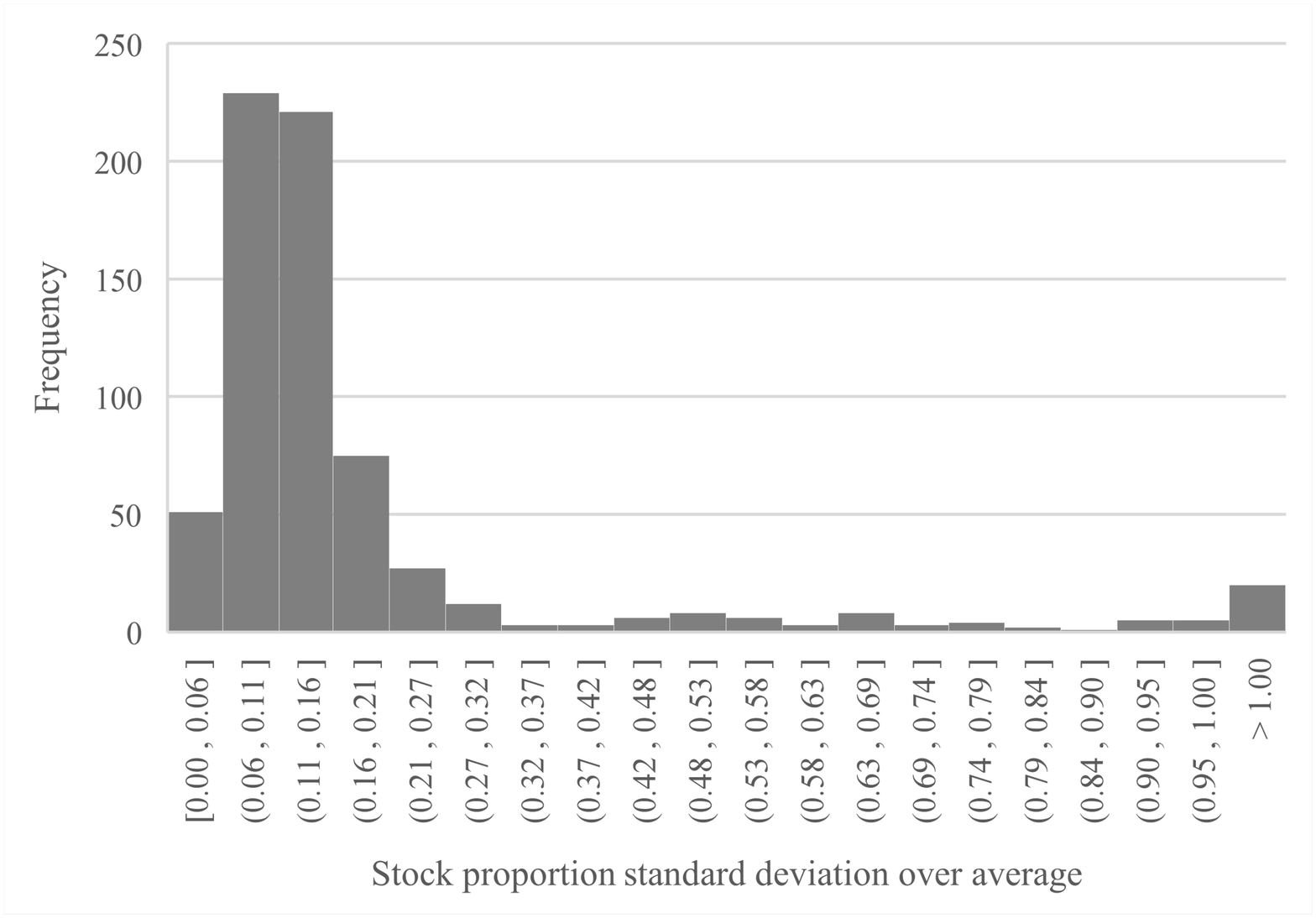}}
	\noindent\caption{}
	{\bf Histogram of fund's stock proportion standard deviation over average.} 
	
	This figure describes the proportion of stock holding within all fund holdings. We consider stock funds with continuous performance data from 2015 to 2017. In this three-year sample, 692 stock funds are present.
	For each fund, we calculate the standard deviation of its stock proportion over its average stock holding proportion from 2015-2017 reports. Then we plot histogram over all funds.  \label{fig:2}
\end{figure}

Figure \ref{fig:2} provides a histogram of fund's stock holding proportion standard deviation over average. 
It shows the size of stock proportion variation compared with the average stock proportion.
More than 70\% funds adjust their stock holding proportion no more than 30\% of their average stock proportion. 
More than 60\% funds adjust their stock holding 
proportion from 10\% to 30\% of their average stock proportion. 
They never deviate too far from their average level of stock holding proportion. 

\bigskip
\centerline{\bf [Place Figure~\ref{fig:2} about here]}
\bigskip

\begin{figure}[!htb]
	\centerline{\includegraphics[width=7in]{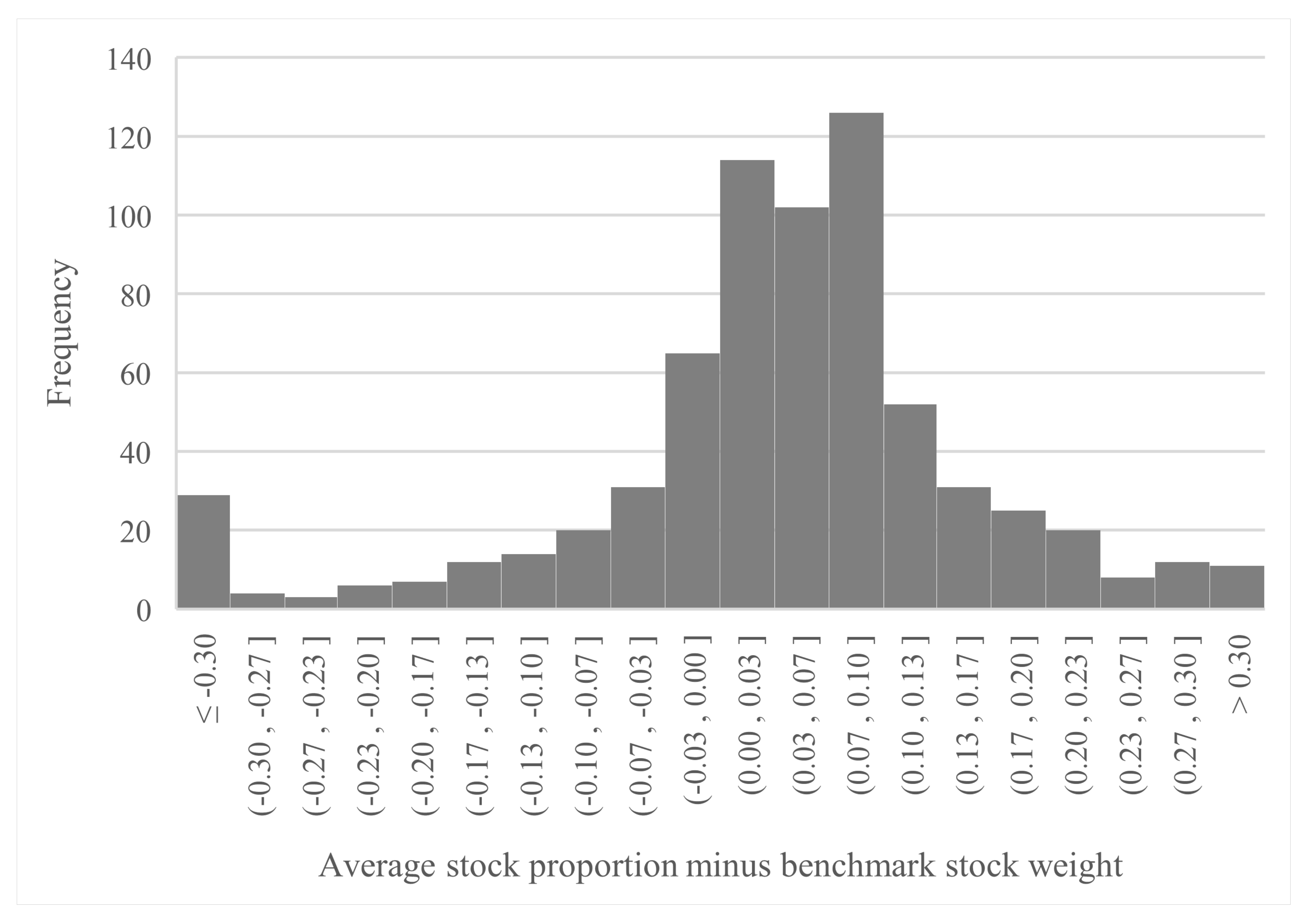}}
	\noindent\caption{}
	{\bf Histogram of fund's average stock holding proportion minus benchmark stock weight.} 
	
	This figure describes the difference between actual stock holding proportion and stock weight in benchmark. We consider stock funds with continuous performance data from 2015 to 2017. In this three-year sample, 692 stock funds are present.
	For each fund, we calculate the average stock holding proportion minus stock weight in its benchmark from 2015-2017 reports. Then we plot histogram over all funds.  \label{fig:3}
\end{figure}

Figure \ref{fig:3} provides a histogram of fund's average stock holding proportion minus benchmark stock weight.
It compares actual stock holding with benchmark to see whether funds follow benchmark in their holdings.
More than 60\% stock funds hold stock with a proportion greater than benchmark stock weight. 
More than 5\% stock funds hold stock with a proportion 20\% greater than benchmark stock weight. 
Most stock funds show the tendency to hold more stocks than indicated in their benchmarks.

\bigskip
\centerline{\bf [Place Figure~\ref{fig:3} about here]}
\bigskip

In China, stock holding details are available at semi-annual frequency. We download semi-annual report on stocks held by stock funds from Wind. 
Table \ref{table:1} lists summary statistics of stock holdings semi-annually for actively managed stock mutual funds in China. We exclude stock funds with severe mismatch between fund and benchmark. 
We can see that the number of such funds increases from around a hundred to around a thousand from 2013 to 2017. 
From 2003 to 2012, the number of such funds is small, especially before 2005. 
Average size is the average capital invested in stocks per fund. We can see that average size of stock investment per fund is greatest in 2007. Then it gradually decreases and the average stock investment per fund in 2017 is only half of that in 2003. 
Average number of stocks is the average number of stocks per fund. It shows how much diversification on average the stock funds hold in that period. We can see that the amount of diversification stock funds need is very stable across time from 2003 to 2017, which implies that fund does not seek more diversification. 
In \citet{Kacperczyk2005}, it shows that in US from 1984 to 1999, funds with less diversification in industries generate better excess return. This might explain why the level of diversification does not change much for stock mutual funds in China from 2005 to 2017. 

\bigskip
\centerline{\bf [Place Table~\ref{table:1} about here]}
\bigskip

As for the data source, we download monthly stock data from Wind including close price and stock industry. We also download benchmark index constituents from Wind, which is a commonly used Chinese finance data provider. For index whose constituents are not included in Wind, we download the list of stock components and use market capitalization as weight. Most index constituents are available from Wind.
We download fund's adjusted net asset value from Csmar. Csmar provides well-organized finance data which is convenient to use for academic purposes. 

\subsection{Benchmark Check}
\label{sec:Data_Check}

From section \ref{sec:Data_Summary}, we can see that we use semi-annual stock holdings details for calculating performance measures from Brinson model. 
In Brinson model (\citet{Brinson1995}), it is assumed that
benchmark is appropriate for fund. We calibrate the fund's ability by comparing it with benchmark performance. 

Comparing fund holdings with self-reported benchmark of stock funds is natural and straightforward. The pre-requirement is that self-reported benchmark are well specified and reflect the investment protocol of funds. For example, actual stock holding proportion fluctuates around benchmark-reported stock weight over time. Industries covered in fund holdings are similar to those included in benchmark index constituents. For some funds, benchmark does not reflect its investment strategy. For example, benchmark is interest rate but fund holdings cover lots of stocks. In some years when benchmark significantly under-performs the market, fund holdings cover absolutely different industries from benchmark. 

\begin{figure}[!htb]
	\centerline{\includegraphics[width=7in]{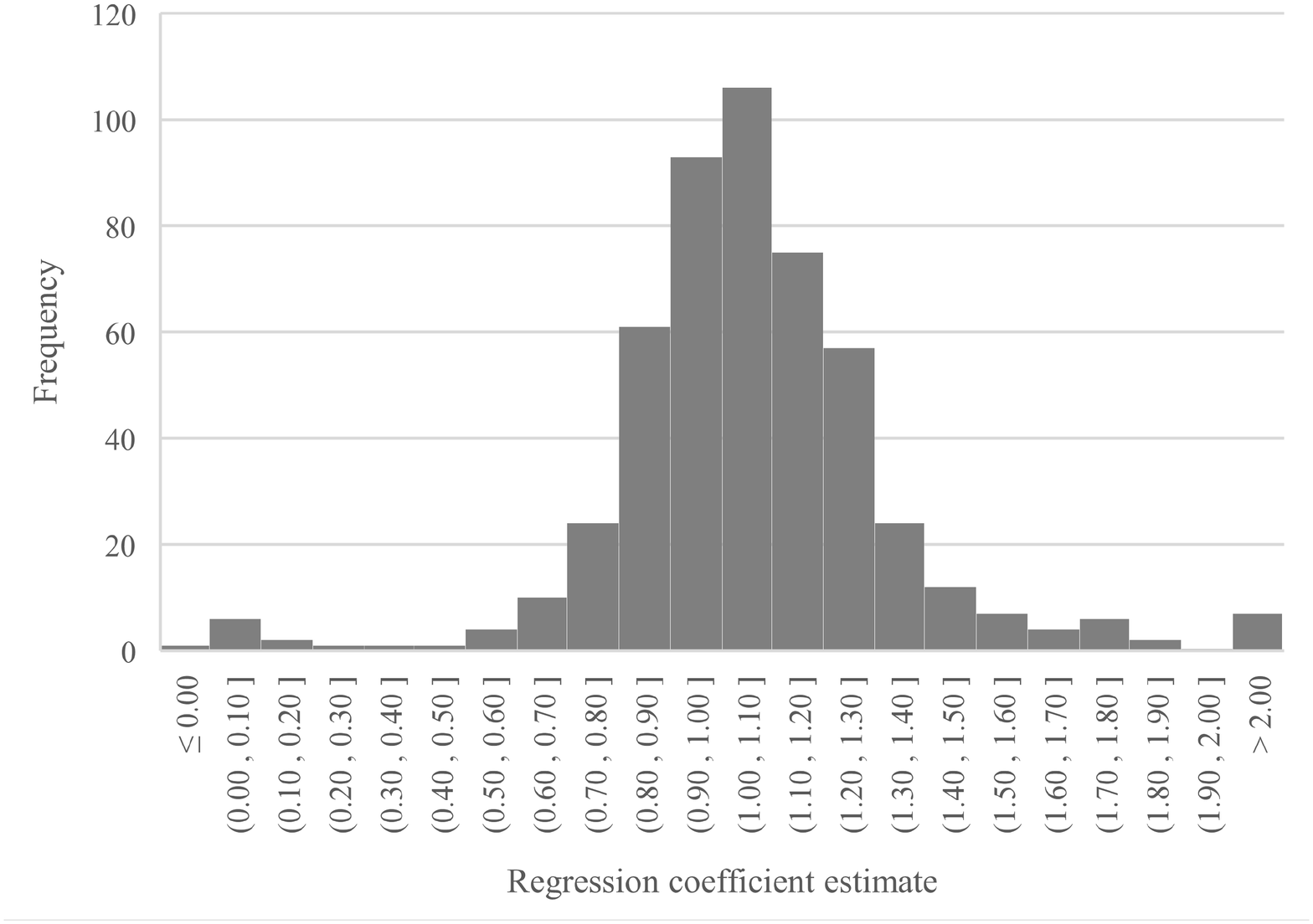}}
	\noindent\caption{}
	{\bf Histogram of regression coefficient $\hat{\beta}_d$ from model (\ref{eq:benchmark_check}).} 
	
	This figure describes whether benchmarks are appropriate for funds. We consider stock funds with continuous performance data from 2013 to 2017. In this five-year sample, 487 stock funds are present.
	For each fund, we run regression in model (\ref{eq:benchmark_check}) to estimate coefficient $\hat{\beta}_d$. Then we plot histogram over all funds.  \label{fig:4}
\end{figure}

First, we consider funds with continuous performance from 2013 to 2017. In this sample, we apply model (\ref{eq:benchmark_check}) to each fund 
and test whether the coefficient is one. We use monthly data to fit the regression model.
In figure \ref{fig:4}, we show the histogram of coefficient estimates for all funds. We can see that 50\%-60\% funds are of regression coefficient greater than 1. About 90\% of fund's regression coefficient lies between 0.7 to 1.4.

\bigskip
\centerline{\bf [Place Figure~\ref{fig:4} about here]}
\bigskip

\begin{figure}[!htb]
	\centerline{\includegraphics[width=7in]{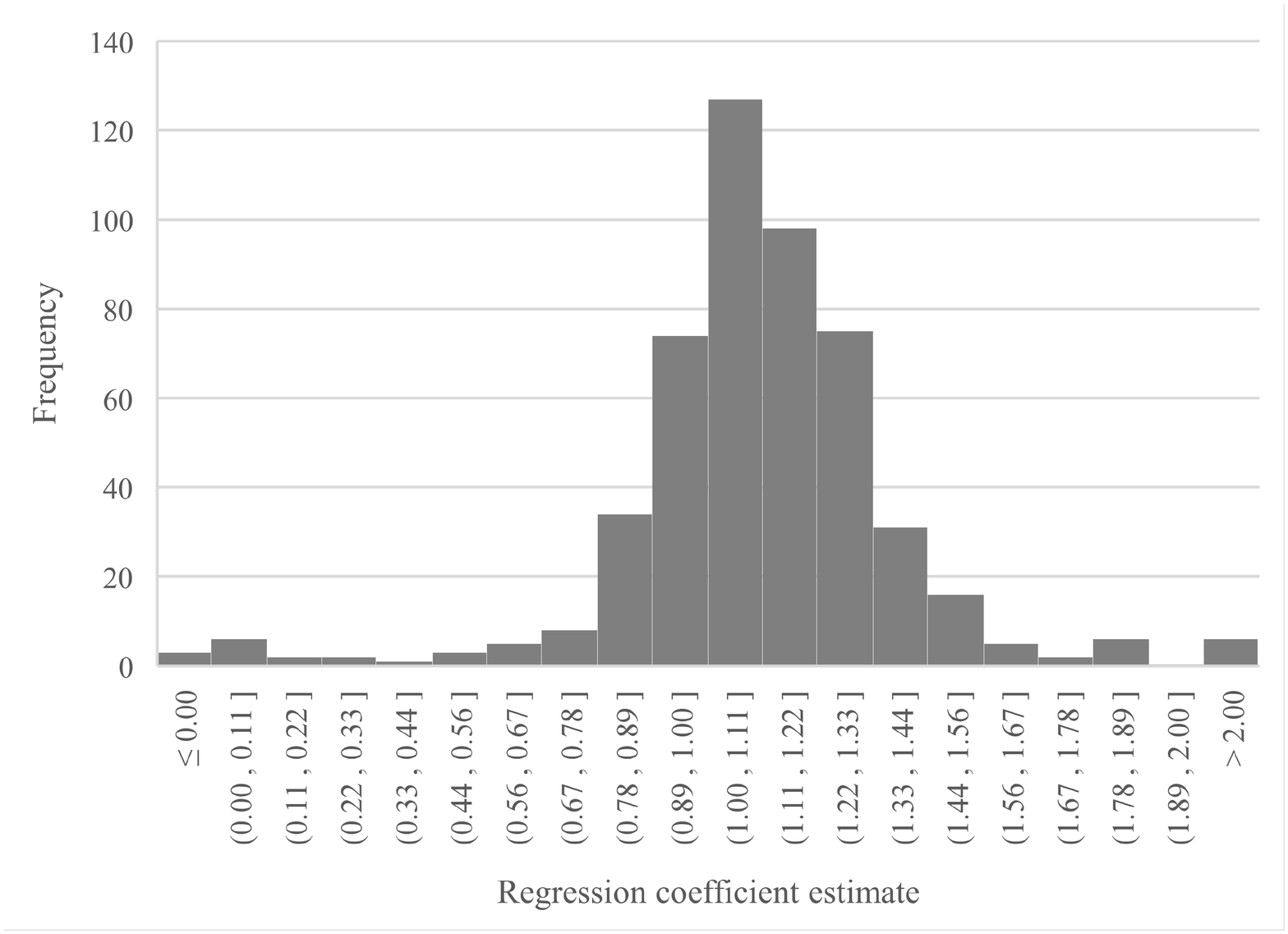}}
	\noindent\caption{}
	{\bf Histogram of regression coefficient $\hat{\beta}_d$ from model (\ref{eq:benchmark_reg1}).} 
	
	This figure describes whether benchmarks are appropriate for funds. We consider stock funds with continuous performance data from 2013 to 2017. In this five-year sample, 487 stock funds are present.
	For each fund, we run regression in model (\ref{eq:benchmark_reg1}) to estimate coefficient $\hat{\beta}_d$. Then we plot histogram over all funds.  \label{fig:8}
\end{figure}

Second, in model \ref{eq:benchmark_reg1}, we add size factor and value factor to the regression and test whether the coefficient is one. As demonstrated in figure \ref{fig:8}, about 70\% of funds are of regression coefficient greater than 1. After adjusting for the effect of size and value factors, the effect of benchmark return upon fund's return is farther away from one on average. Most funds take additional risk other than benchmark. 
The additional risk is not explainable by size or value factors.  

\bigskip
\centerline{\bf [Place Figure~\ref{fig:8} about here]}
\bigskip

In table \ref{table:3}, it shows the comparison between coefficient estimates under model \ref{eq:benchmark_check} and model \ref{eq:benchmark_reg1}. We can see that under model \ref{eq:benchmark_reg1} more benchmarks are tested to be improper for the corresponding stock funds. After considering size and value factors, the mismatch between funds and benchmarks is more severe. Under model \ref{eq:benchmark_check}, 28\% of funds show benchmark mismatch when we consider 95\% confidence interval. In this sample, 72\% of funds do not show benchmark mismatch from the viewpoint of regression model. 

\bigskip
\centerline{\bf [Place Table~\ref{table:3} about here]}
\bigskip

\begin{figure}[!htb]
	\centerline{\includegraphics[width=7in]{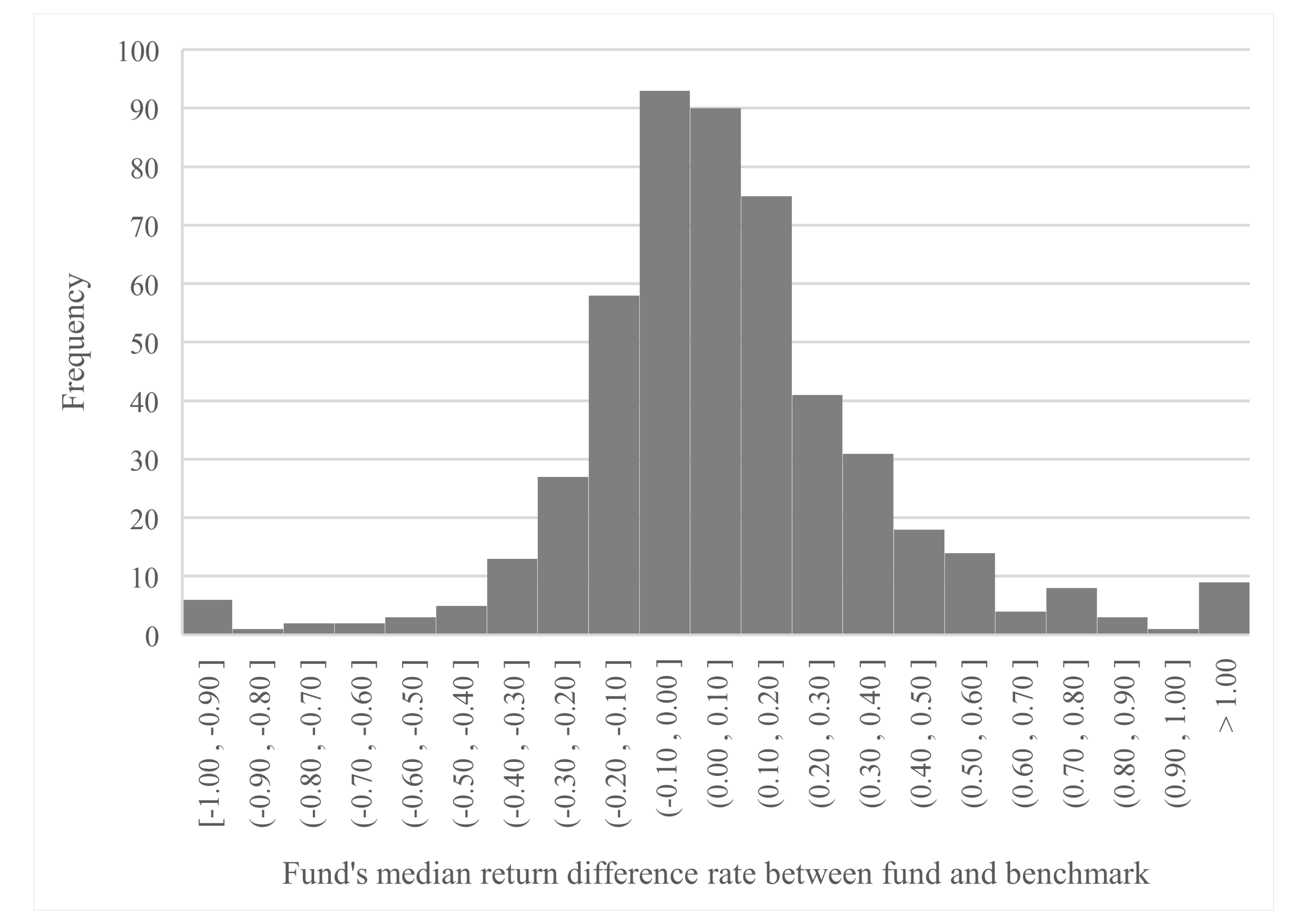}}
	\noindent\caption{}
	{\bf Histogram of fund's median return difference between fund and benchmark.} 
	
	This figure describes the difference between fund return and benchmark return. We consider stock funds with continuous performance data from 2013 to 2017. In this five-year sample, 487 stock funds are present.
	Denote fund return by $r_t$ and the corresponding benchmark return by $r_{dt}$.
	We calculate the median of $(r_t-r_{dt})/r_{dt}$, $t=1,2,\cdots,T$ for each fund and plot the histogram over all stock funds in five-year sample.  
	\label{fig:5}
\end{figure}

Finally, we calculate fund's return and benchmark return in every semi-annual period. We also calculate the median difference between fund's return and benchmark return. In figure \ref{fig:5}, we show the histogram of all the median differences. We can see that the median difference of all funds is symmetric about zero. It implies that the returns of fund and benchmark are very close on average and that benchmark is a good indicator of fund's return on average.

\bigskip
\centerline{\bf [Place Figure~\ref{fig:5} about here]}
\bigskip

In this section, we conclude that for 70\% of stock funds, their benchmarks are proper and reflect their investment protocol. On average, actual return of stock fund is close to its benchmark return. 

\section{Performance Measures Summary}
\label{sec:Performance_Measures_Summary}

We examine funds with continuous performance from 2015 to 2017.
In this sample, we calculate performance measures: industry allocation, within-industry selection and interaction term at every semi-annual report. In formulas (\ref{eq:ia}), (\ref{eq:wiss}) and (\ref{eq:it}), we use six months before every semi-annual report to calculate the return in these formulas.
We also calculate asset allocation at every quarterly report.
In formula (\ref{eq:aa}), we use three months after every quarterly report to calculate the return in formula. 

For every fund, we calculate average industry allocation, within-industry selection, interaction term and asset allocation. We also calculate standard deviation of each performance measure.
We compute the proportion of fund whose performance measure is positive.
We use t test to see whether performance measure is significantly positive or not.
We use test statistic $\sqrt{n}\bar{x}/\hat{\sigma}_x$ and limiting distribution $t_{n-2}$ to conduct the hypothesis test.
For performance measures based upon semi-annual reports, $n=6$. For performance measures based upon quarterly reports, $n=12$.

\bigskip
\centerline{\bf [Place Table~\ref{table:4} about here]}
\bigskip

We summarize above results in table \ref{table:4}. We can see that 72\% of funds show positive within-industry selection and 40\% of funds show significantly positive within-industry selection. Only 16\% funds show positive asset allocation and 3\$ show significantly positive asset allocation. 25\% funds show positive industry allocation and 8\% funds show significantly positive industry allocation. Most funds show positive stock selection ability but not asset allocation ability.

Recall that interaction term indicates whether funds simultaneously over-weight superior industry and superior stocks within industry, or simultaneously under-weight superior industry and superior stocks within industry. 
In our results, we can see that 96\% funds show positive interaction term and 55\% funds show significantly positive interaction term. For most funds, they either exhibit both positive industry allocation and within-industry selection, or exhibit both negative industry allocation and within-industry selection. 

\begin{figure}[!htb]
	\centerline{\includegraphics[width=7in]{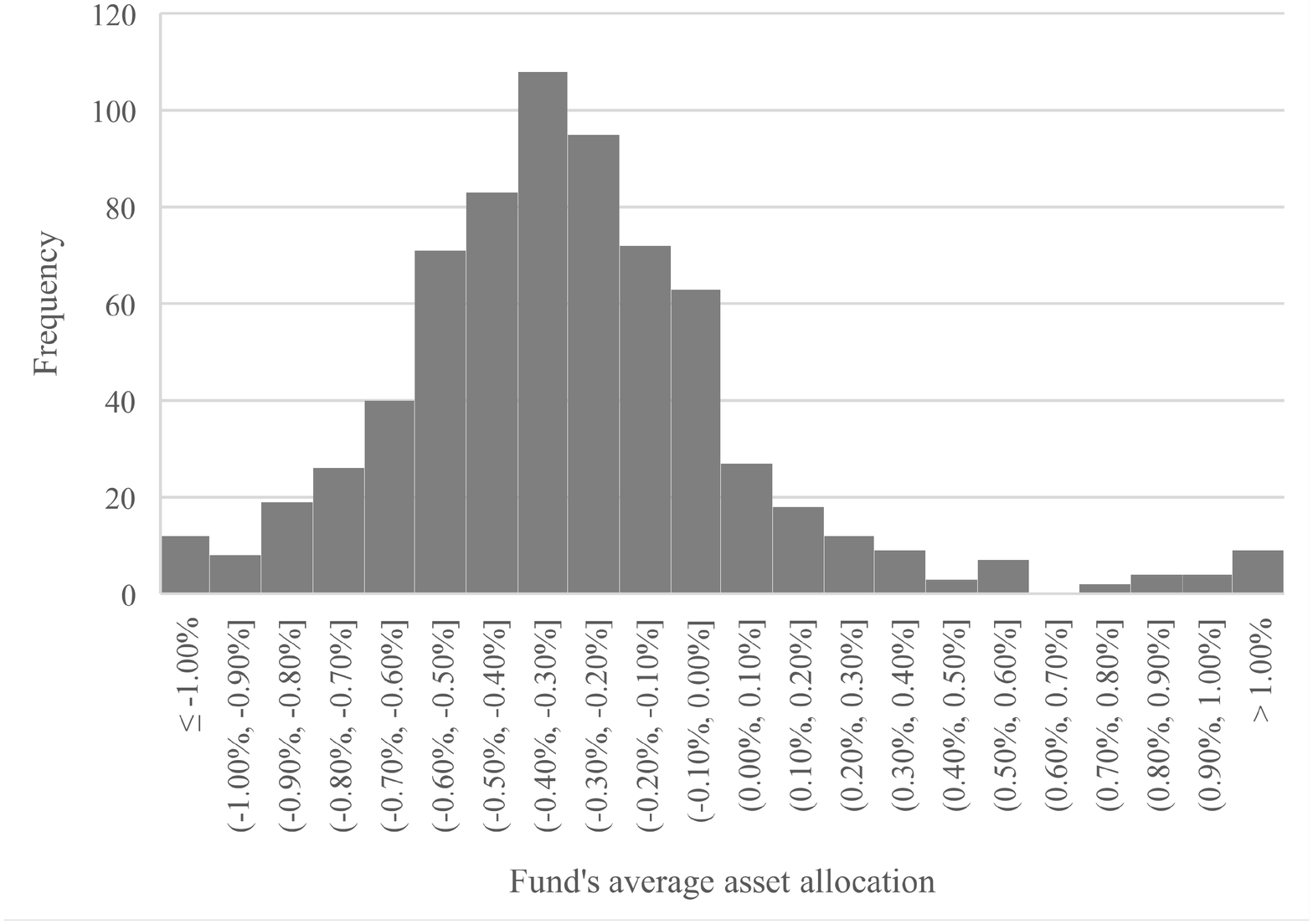}}
	\noindent\caption{}
	{\bf Histogram of fund's average asset allocation.} 
	
	This figure describes the fund's average asset allocation. We consider stock funds with continuous performance data from 2015 to 2017. In this three-year sample, 692 stock funds are present.
	For each fund, we calculate the average asset allocation from 2015-2017 reports. Then we plot histogram over all funds.  \label{fig:7}
\end{figure}

\bigskip
\centerline{\bf [Place Figure~\ref{fig:7} about here]}
\bigskip

Next we see from figures the rough distribution of these performance measures for all the stock funds.
First, in figure \ref{fig:7}, we can see that average asset allocation is around -0.3\% for all funds. Only a small proportion of funds show positive average asset allocation. Still less than 10 funds show average asset allocation greater than 1\%. Less than 20 funds show average asset allocation greater than 0.7\%. Although most funds show negative asset allocation, about 10 funds show very good asset allocation ability. 

\begin{figure}[!htb]
	\centerline{\includegraphics[width=7in]{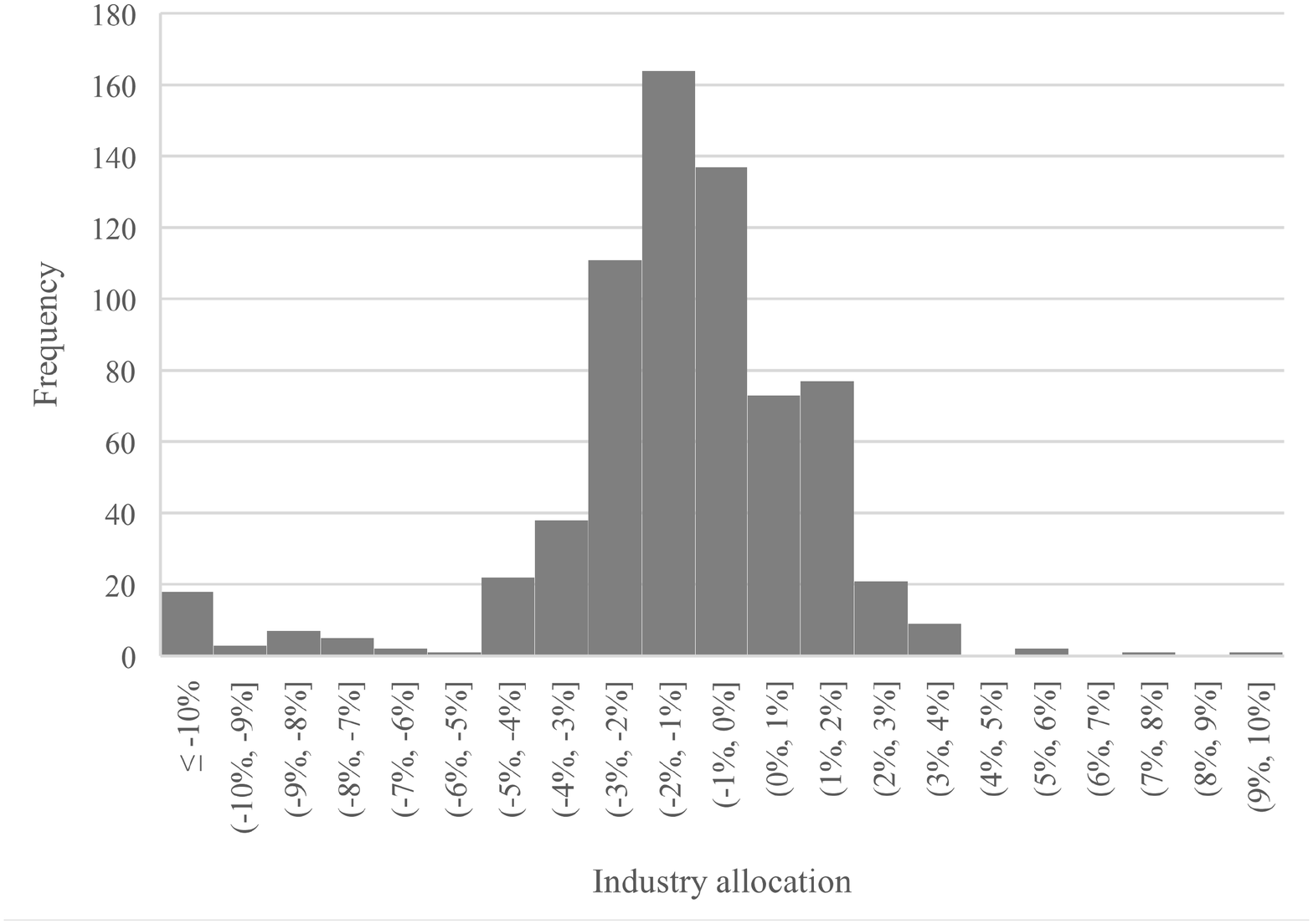}}
	\noindent\caption{}
	{\bf Histogram of fund's average industry allocation.} 
	
	This figure describes the fund's average industry allocation. We consider stock funds with continuous performance data from 2015 to 2017. In this three-year sample, 692 stock funds are present.
	For each fund, we calculate the average industry allocation from 2015-2017 semi-annual reports. Then we plot histogram over all funds. \label{fig:12}
\end{figure}

\bigskip
\centerline{\bf [Place Figure~\ref{fig:12} about here]}
\bigskip

Second, from figure \ref{fig:12}, we can see that most funds show negative industry allocation, but not as many as funds with negative asset allocation. On average industry allocation is around -1\% and there are about 20 funds with industry allocation less than -10\%. 

\begin{figure}[!htb]
	\centerline{\includegraphics[width=7in]{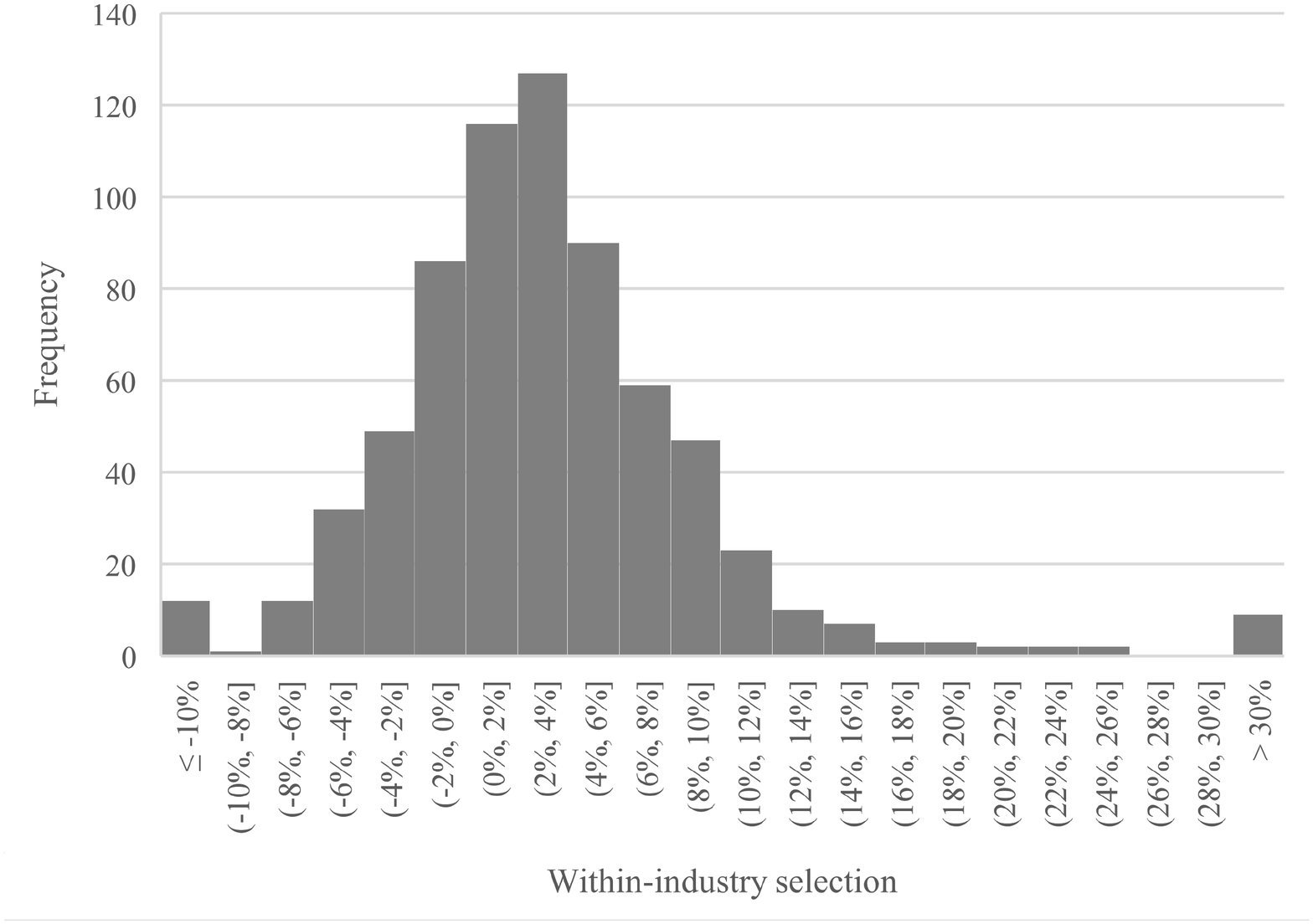}}
	\noindent\caption{}
	{\bf Histogram of fund's average within-industry selection.} 
	
	This figure describes the fund's average within-industry selection. We consider stock funds with continuous performance data from 2015 to 2017. In this three-year sample, 692 stock funds are present.
	For each fund, we calculate the average within-industry selection from 2015-2017 semi-annual reports. Then we plot histogram over all funds. \label{fig:13}
\end{figure}

\bigskip
\centerline{\bf [Place Figure~\ref{fig:13} about here]}
\bigskip

Third, in figure \ref{fig:13}, we show that most stock funds exhibit positive within-industry selection. Average within-industry selection is around 2\%. There are about 8 funds with within-industry selection greater than 30\%. More than 40 funds show within-industry selection greater than 10\%. Only about 10 funds show within-industry selection less than -10\%. 

\begin{figure}[!htb]
	\centerline{\includegraphics[width=7in]{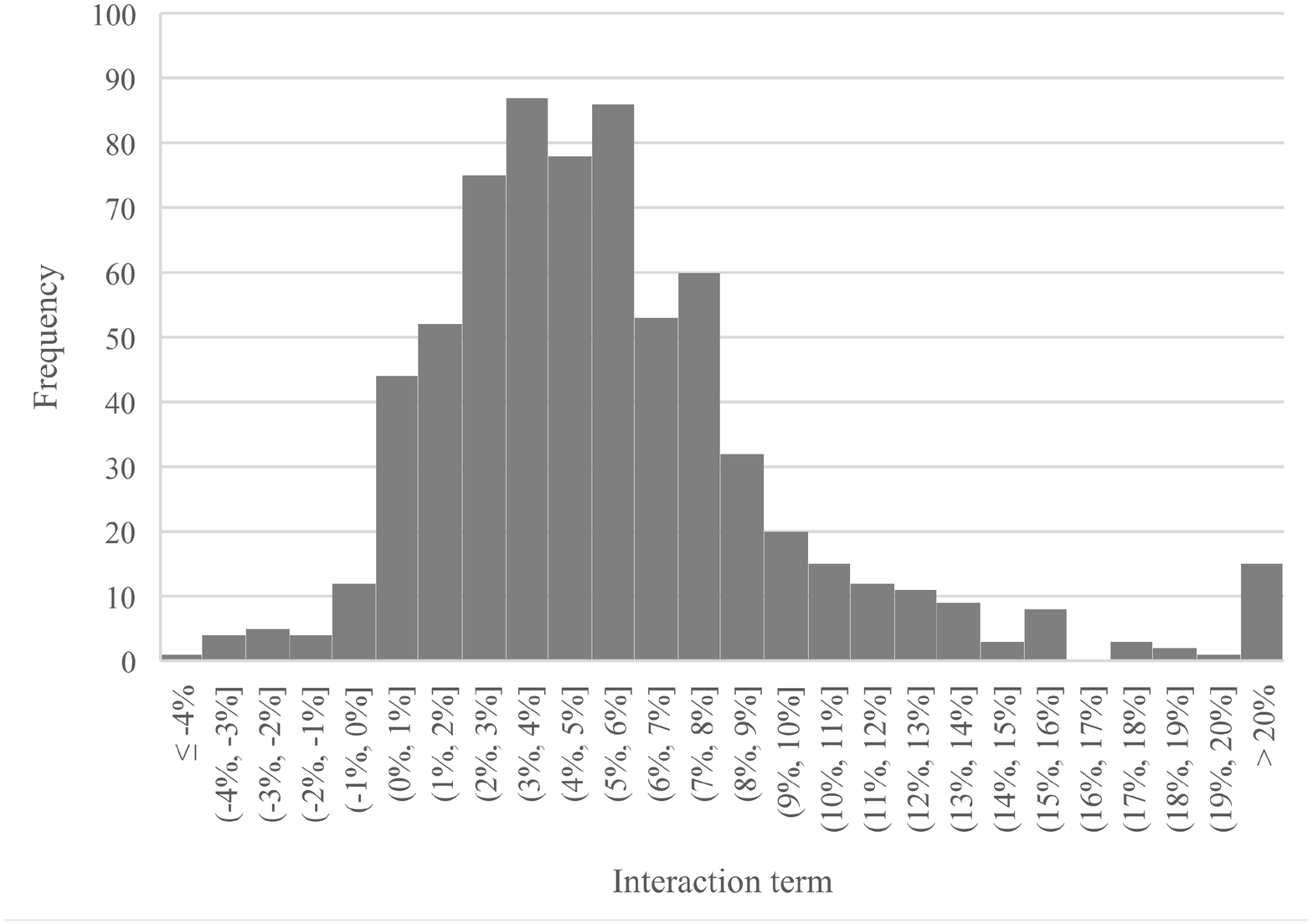}}
	\noindent\caption{}
	{\bf Histogram of fund's average interaction term.} 
	
	This figure describes the fund's average interaction term. We consider stock funds with continuous performance data from 2015 to 2017. In this three-year sample, 692 stock funds are present.
	For each fund, we calculate the average interaction term from 2015-2017 semi-annual reports. Then we plot histogram over all funds. \label{fig:14}
\end{figure}

\bigskip
\centerline{\bf [Place Figure~\ref{fig:14} about here]}
\bigskip

Finally, figure \ref{fig:14} shows histogram of fund's average interaction term. We can see that most funds show positive interaction term. Only 4\% of funds show negative interaction term. More than 15 funds show interaction term greater than 20\%. Funds tend to simultaneously over-weight superior industry and superior stocks within industry, or simultaneously under-weight superior industry and superior stocks within industry.

\begin{figure}[!htb]
	\centerline{\includegraphics[width=7in]{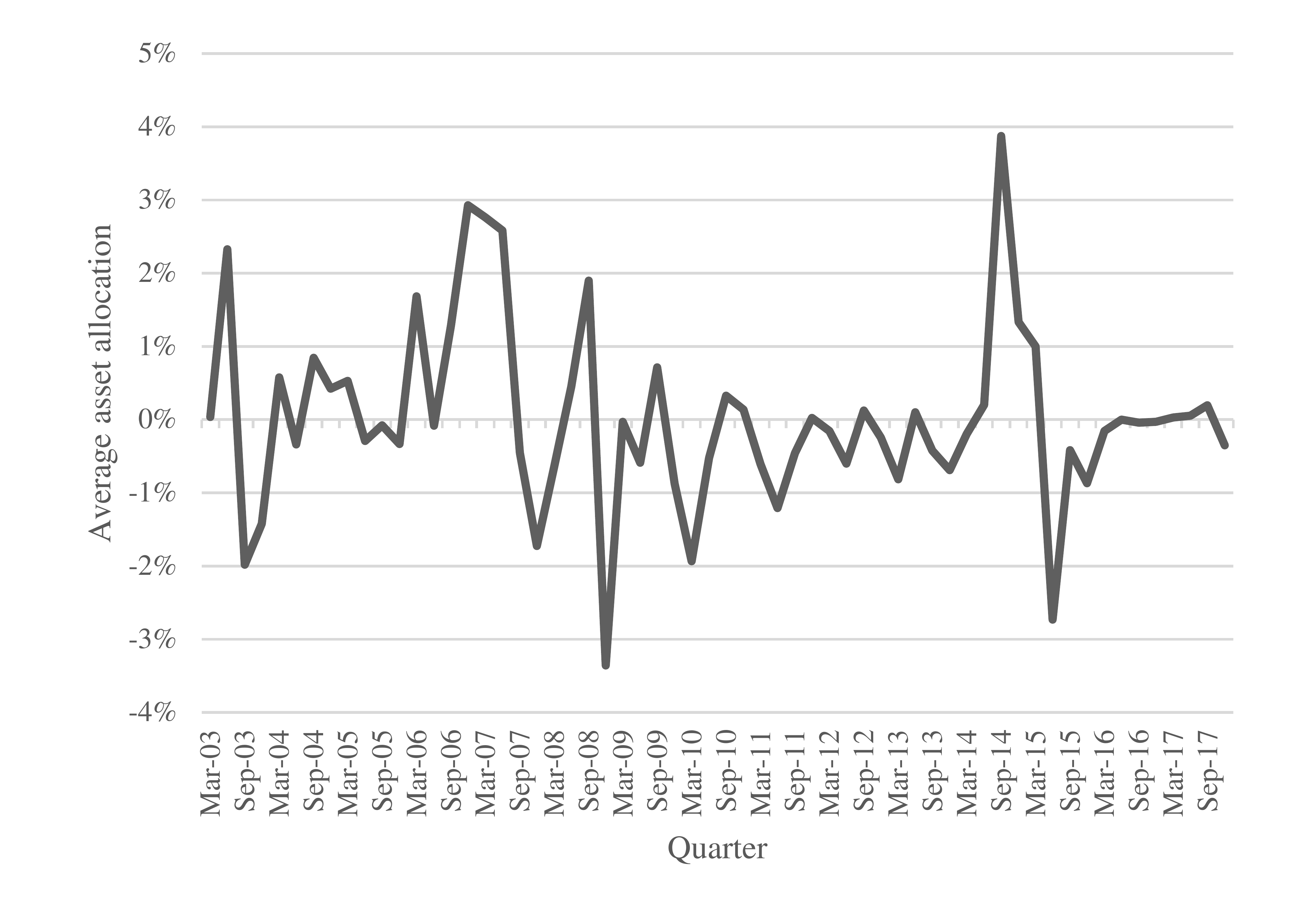}}
	\noindent\caption{}
	{\bf Graph of average asset allocation ability.} 
	
	This figure describes the average of asset allocation by quarter. In each quarterly report, we calculate the average asset allocation of actively managed stock mutual funds. Then we plot a graph of average asset allocation over quarters.   \label{fig:6}
\end{figure}

\bigskip
\centerline{\bf [Place Figure~\ref{fig:6} about here]}
\bigskip

Now we look at results for all stock funds from 2003 to 2017. We consider average performance measures at every semi-annual report or quarterly report.
Figure \ref{fig:6} shows average asset allocation ability of all stock funds at every quarterly report. 
We can see that over time average asset allocation fluctuates around 0\%. In years of great change, such as years 2007-2008 and years 2014-2015, average asset allocation shows drastic changes as well. At time of crisis, on average, funds do not see the crisis coming and fail to make corresponding asset allocation, which drags average asset allocation down to -3\%.  
In years 2016-2017, average asset allocation is very stable over time with slight decrease in late 2017.

\begin{figure}[!htb]
	\centerline{\includegraphics[width=7in]{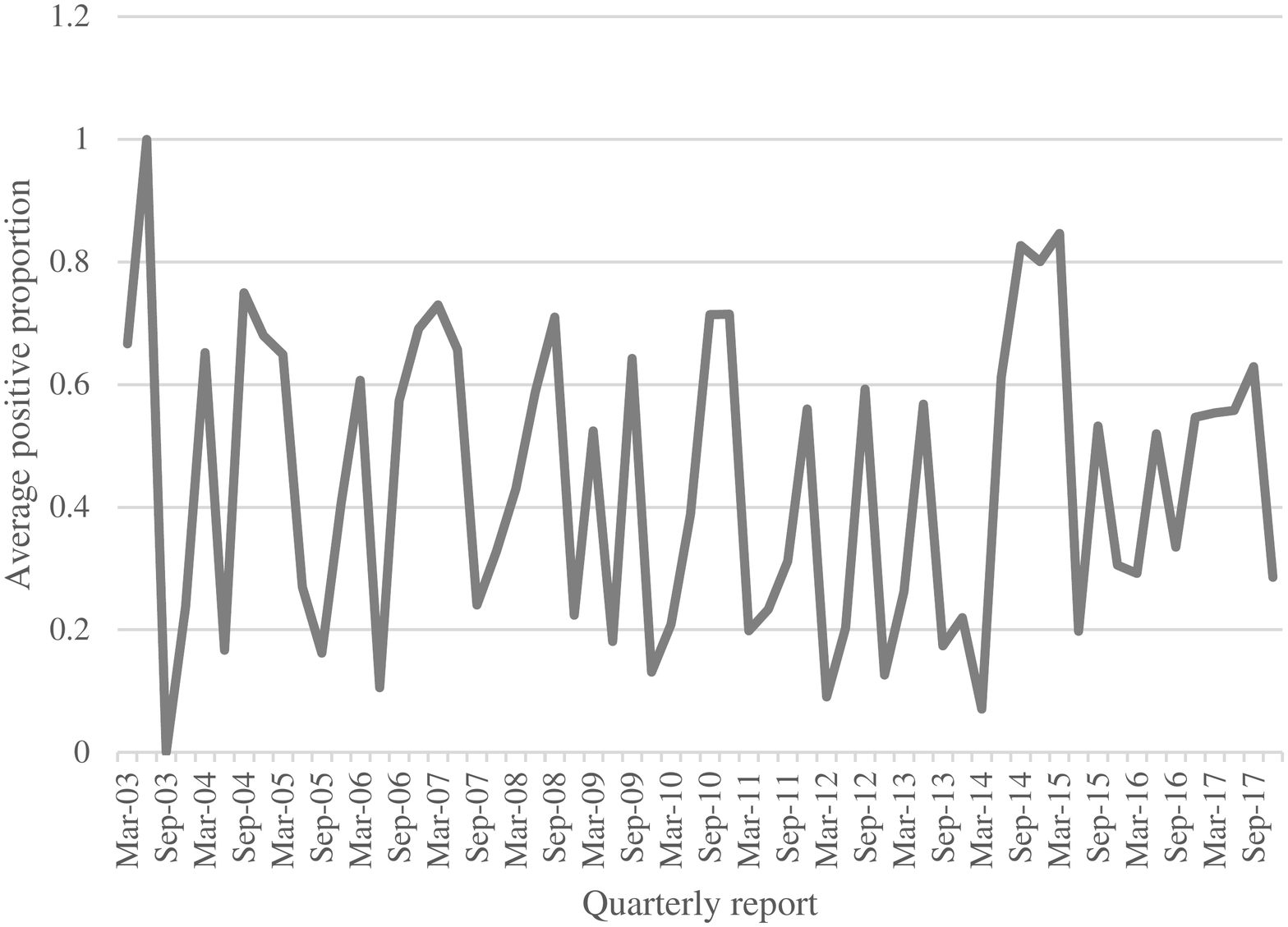}}
	\noindent\caption{}
	{\bf Graph of average positive proportion of fund's asset allocation.} 
	
	This figure describes average proportion of positive fund's asset allocation at quarterly reports. In each quarterly report, we consider all stock funds in it, calculate their asset allocation and the proportion of funds whose asset allocation is positive. Then we plot average positive proportion over all quarterly reports.  \label{fig:9}
\end{figure}

\bigskip
\centerline{\bf [Place Figure~\ref{fig:9} about here]}
\bigskip

In figure \ref{fig:9}, we can see that the proportion of funds with positive asset allocation at every quarterly report lies between 0.2 and 0.7 most of the time and fluctuates around 0.4. From late 2014 to early 2015, positive proportion rises to over 0.8, which implies that most funds show positive asset allocation in bullish market. 
However, in bearish market which comes after bulls, positive proportion decreases to 0.2, which is not the lowest proportion over all times. That is, in bearish market, a normal proportion of funds show positive asset allocation, but in bullish market, an abnormal high proportion of funds show positive asset allocation. 

\begin{figure}[!htb]
	\centerline{\includegraphics[width=7in]{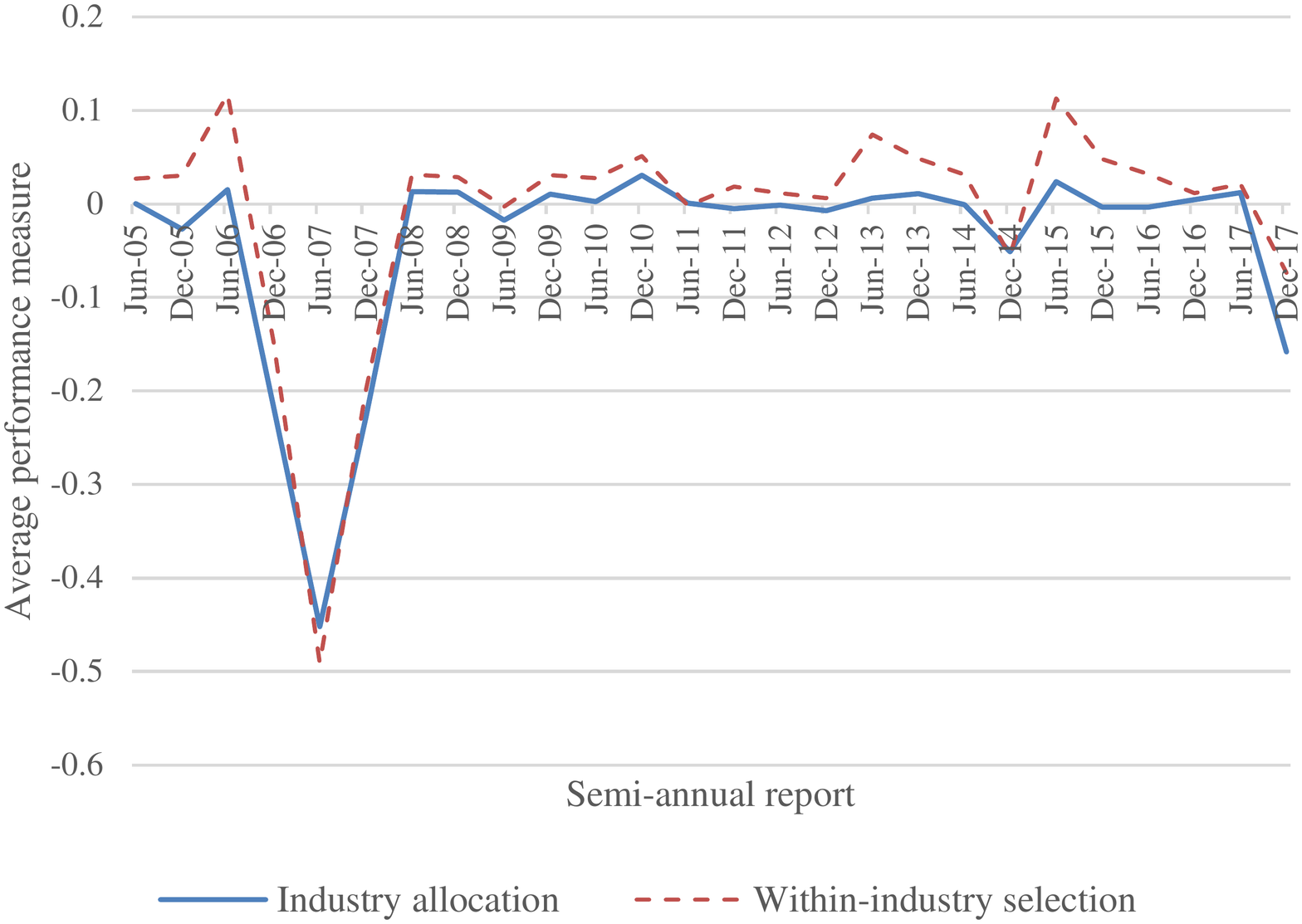}}
	\noindent\caption{}
	{\bf Graph of average industry allocation and within-industry selection.} 
	
	This figure describes the average industry allocation and within-industry selection at every semi-annual report. In each semi-annual report, we calculate the average industry allocation and within-industry selection of actively managed stock mutual funds. Then we plot a graph of average industry allocation and within-industry selection over semi-annual reports.   \label{fig:20}
\end{figure}

\bigskip
\centerline{\bf [Place Figure~\ref{fig:20} about here]}
\bigskip

Here in figure \ref{fig:20}, we show average industry allocation and within-industry selection over time. We can see that they are positive most of the time from late 2008 to early 2017. These two averages decline down to -0.5 in 2007 before the crisis. And they decrease to -0.1 in 2017. We also notice that average industry allocation and within-industry selection move together in similar direction. 

\begin{figure}[!htb]
	\centerline{\includegraphics[width=7in]{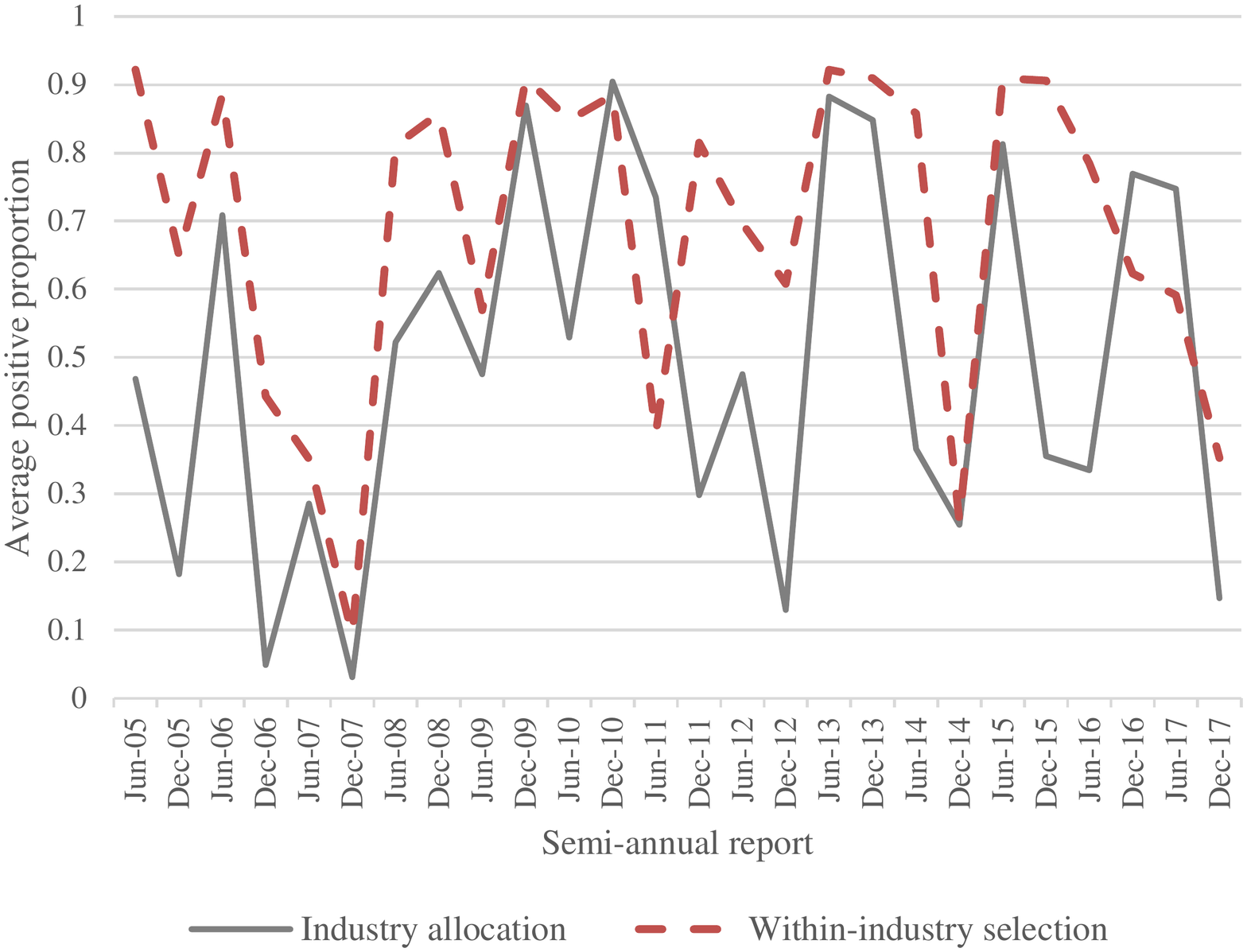}}
	\noindent\caption{}
	{\bf Graph of average positive proportion of fund's industry allocation and within-industry selection.} 
	
	This figure describes average proportion of positive fund's industry allocation and within-industry selection at the semi-annual reports. For each semi-annual report, we consider all stock funds in it, calculate their performance measures and the proportion of funds whose performance measure is positive. Then we plot average positive proportion over all quarterly reports. Here we show results of two performance measures: industry allocation and within-industry selection. \label{fig:10}
\end{figure}

\bigskip
\centerline{\bf [Place Figure~\ref{fig:10} about here]}
\bigskip

Furthermore, figure \ref{fig:10} provides the average positive proportion of industry allocation and within-industry selection at every semi-annual report. Positive proportion of these two measures move together in similar direction. Positive proportion fluctuates between 0.1 and 0.9. In crisis years, average measures tumble to all-time lows but their positive proportions decrease to 0.1 which is relatively normal.

\begin{figure}[!htb]
	\centerline{\includegraphics[width=7in]{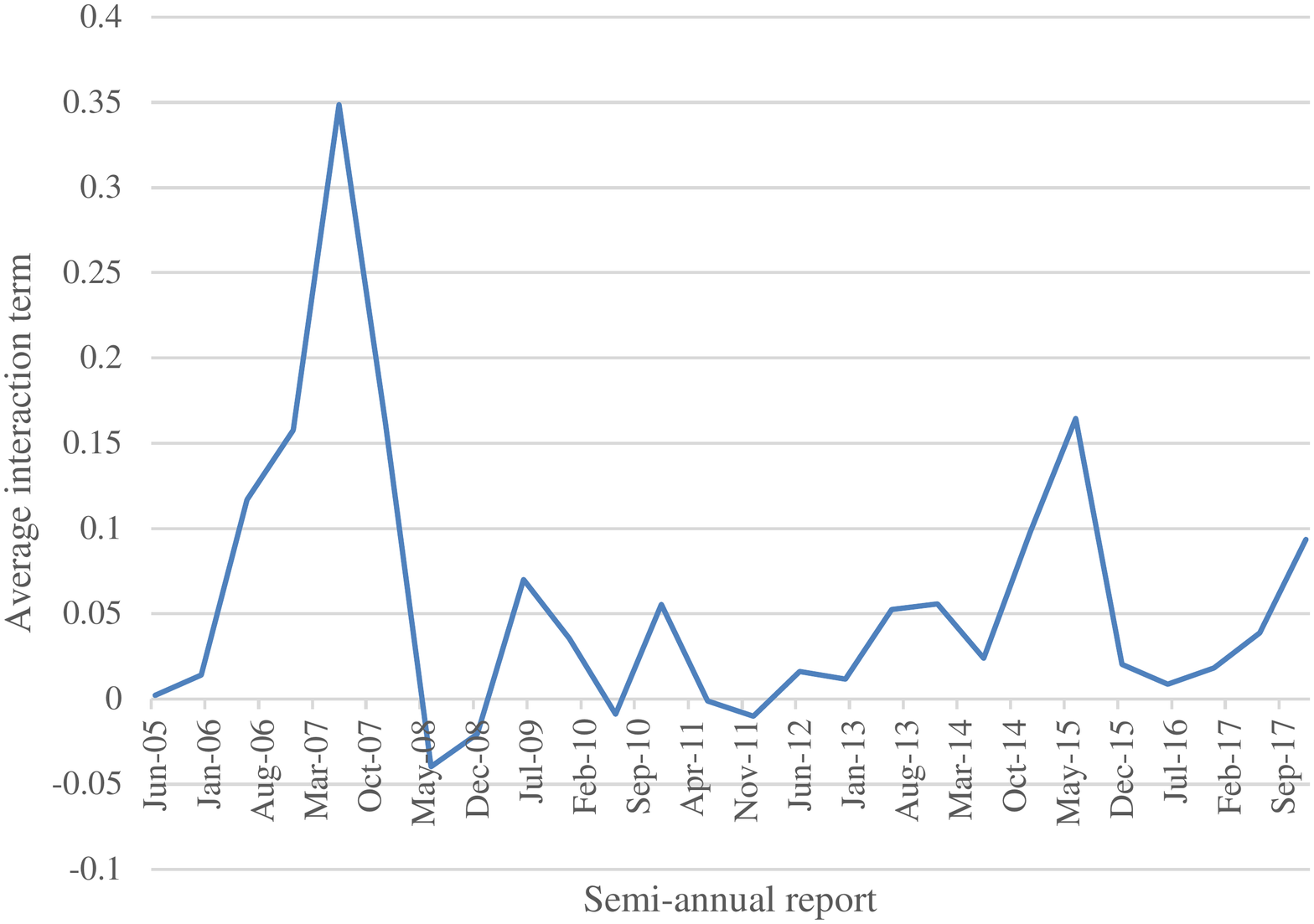}}
	\noindent\caption{}
	{\bf Graph of average interaction term.} 
	
	This figure describes the average interaction term at every semi-annual report. In each semi-annual report, we calculate the average interaction term of actively managed stock mutual funds. Then we plot a graph of average interaction term over semi-annual reports.   \label{fig:21}
\end{figure}

\bigskip
\centerline{\bf [Place Figure~\ref{fig:21} about here]}
\bigskip

Among all the performance measures, interaction term is the most positive one. It implies that fund's industry allocation ability is positively associated with its within-industry selection ability. From figure \ref{fig:21}, we can see that average interaction term is positive in all semi-annual reports except in 2008 and 2011. The lowest average interaction term is -0.05 and the highest average interaction term is 0.35. In either bad times or good ones, interaction term is very positive. In bad times, funds over-weight inferior industry and inferior stocks within industry. In good times, funds over-weight superior industry and superior stocks within industry.

\begin{figure}[!htb]
	\centerline{\includegraphics[width=7in]{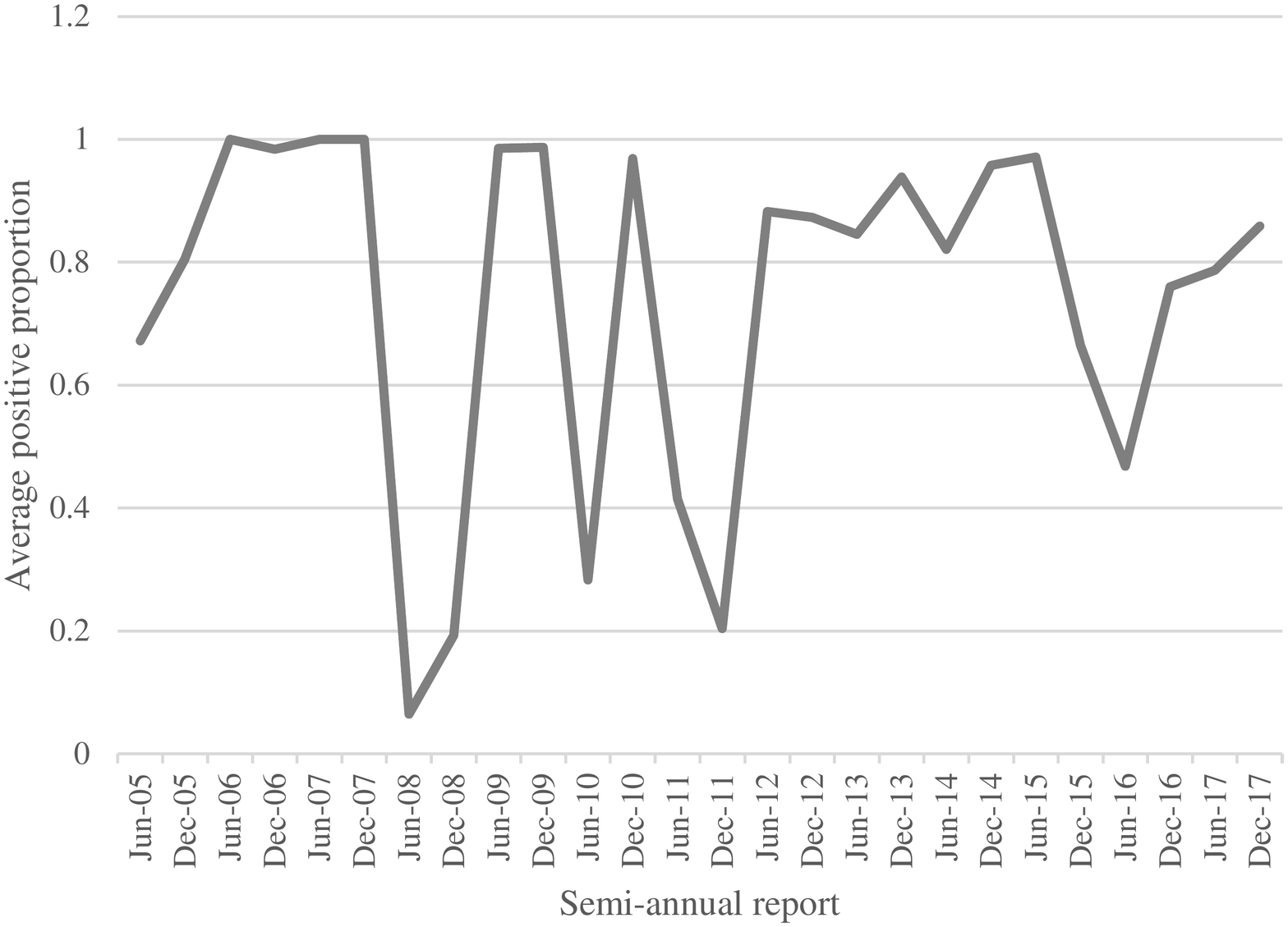}}
	\noindent\caption{}
	{\bf Graph of average positive proportion of fund's interaction term.} 
	
	This figure describes average proportion of positive fund's interaction term at the semi-annual reports. For each semi-annual report, we consider all stock funds in it, calculate their interaction terms and the proportion of funds whose interaction term is positive. Then we plot average positive proportion over all quarterly reports. \label{fig:11}
\end{figure}

\bigskip
\centerline{\bf [Place Figure~\ref{fig:11} about here]}
\bigskip

In figure \ref{fig:11}, we present positive proportion of interaction term over all semi-annual reports. Positive proportion lies between 0.1 and 1 most of the time. In good years with no crisis, positive proportion is very high and close to 1. In crisis years, such as 2008, 2011 and 2015, positive proportion drops to around 0.4. It implies that fund's industry allocation and within-industry selection are more positively associated in good years than in bad years.

In this section, we summarize the performance measures and conclude that most funds show positive within-industry selection but negative asset allocation. Fund's industry allocation and within-industry selection are positively associated and their association is more positive in bulls than in bears. 

\section{Persistence of Performance Measures} \label{sec:Persistence}

In this section, we utilize model in (\ref{eq:persistence_check}) to see whether fund's performance measures are persistent over time or not. We fit model (\ref{eq:persistence_check}) and test the hypothesis in (\ref{eq:persistence_check_hypothesis}) to see whether performance persistence is significant or not. 
We study funds with continuous performance from 2015 to 2017 and we use performance measures from 2015 to 2017 to fit the regression model in (\ref{eq:persistence_check}). Performance measures defined in (\ref{eq:ia}), (\ref{eq:wiss}) and (\ref{eq:it}) use return for six months before each semi-annual report. Asset allocation defined in (\ref{eq:aa}) uses return for three months after each quarterly report.

\begin{figure}[!htb]
	\centerline{\includegraphics[width=7in]{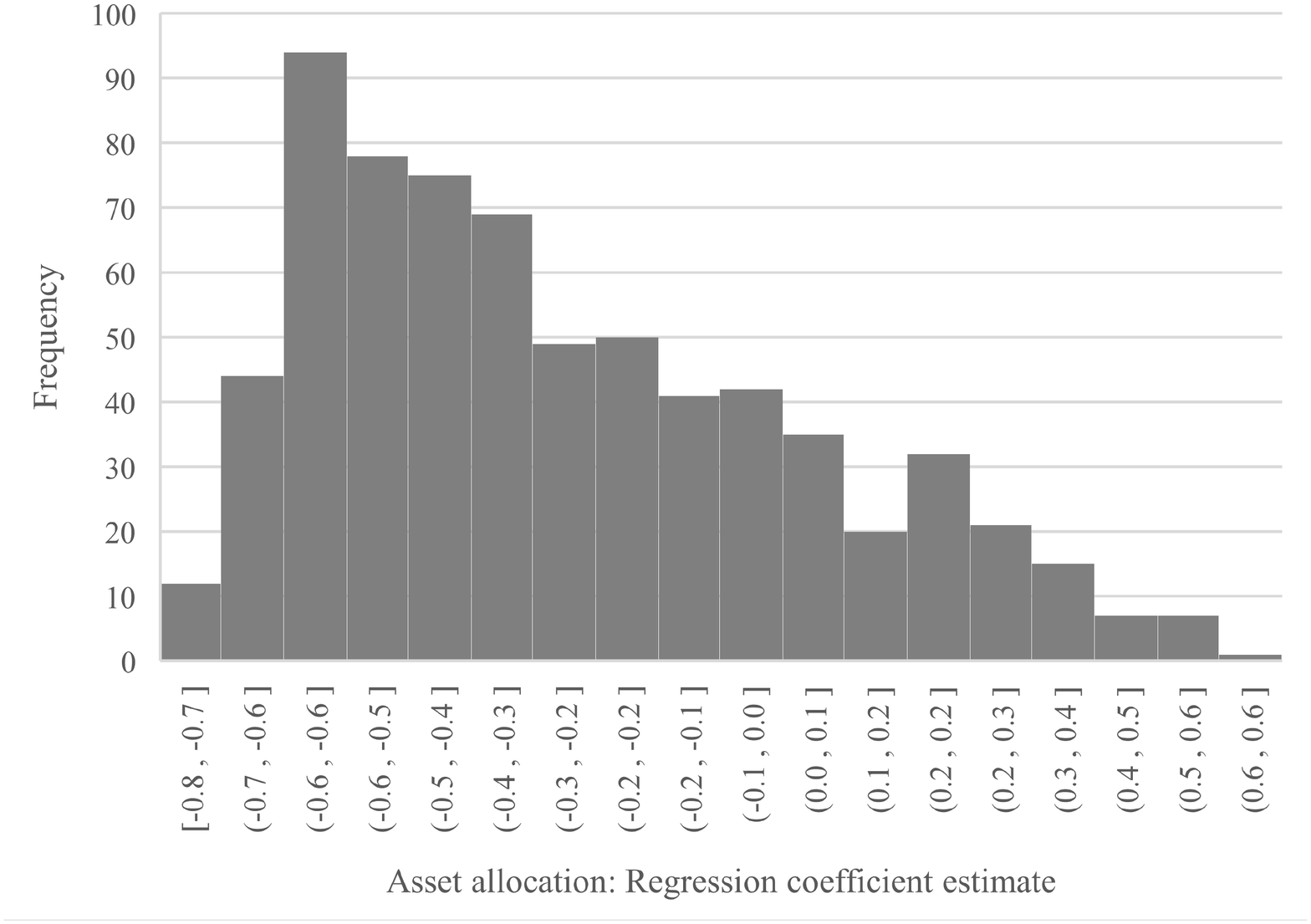}}
	\noindent\caption{}
	{\bf Histogram of fund's asset allocation regression coefficient estimate under model (\ref{eq:persistence_check}).} 
	
	This figure describes the performance persistence of fund's asset allocation. We consider stock funds with continuous performance data from 2015 to 2017. In this three-year sample, 692 stock funds are present.
	For each fund, we calculate the regression coefficient estimate in model (\ref{eq:persistence_check}) using data from 2015-2017 semi-annual reports. Then we plot histogram over all funds. \label{fig:17}
\end{figure}

\bigskip
\centerline{\bf [Place Figure~\ref{fig:17} about here]}
\bigskip

First, in figure \ref{fig:17}, we present regression coefficient estimate of asset allocation. We can see that for most funds, coefficient estimate is negative. That is, for most funds, asset allocation does not show persistence over time. For about 20 funds, coefficient estimate is greater than 0.4, which implies that asset allocation shows positive persistence over time between semi-annual reports. 

\begin{figure}[!htb]
	\centerline{\includegraphics[width=7in]{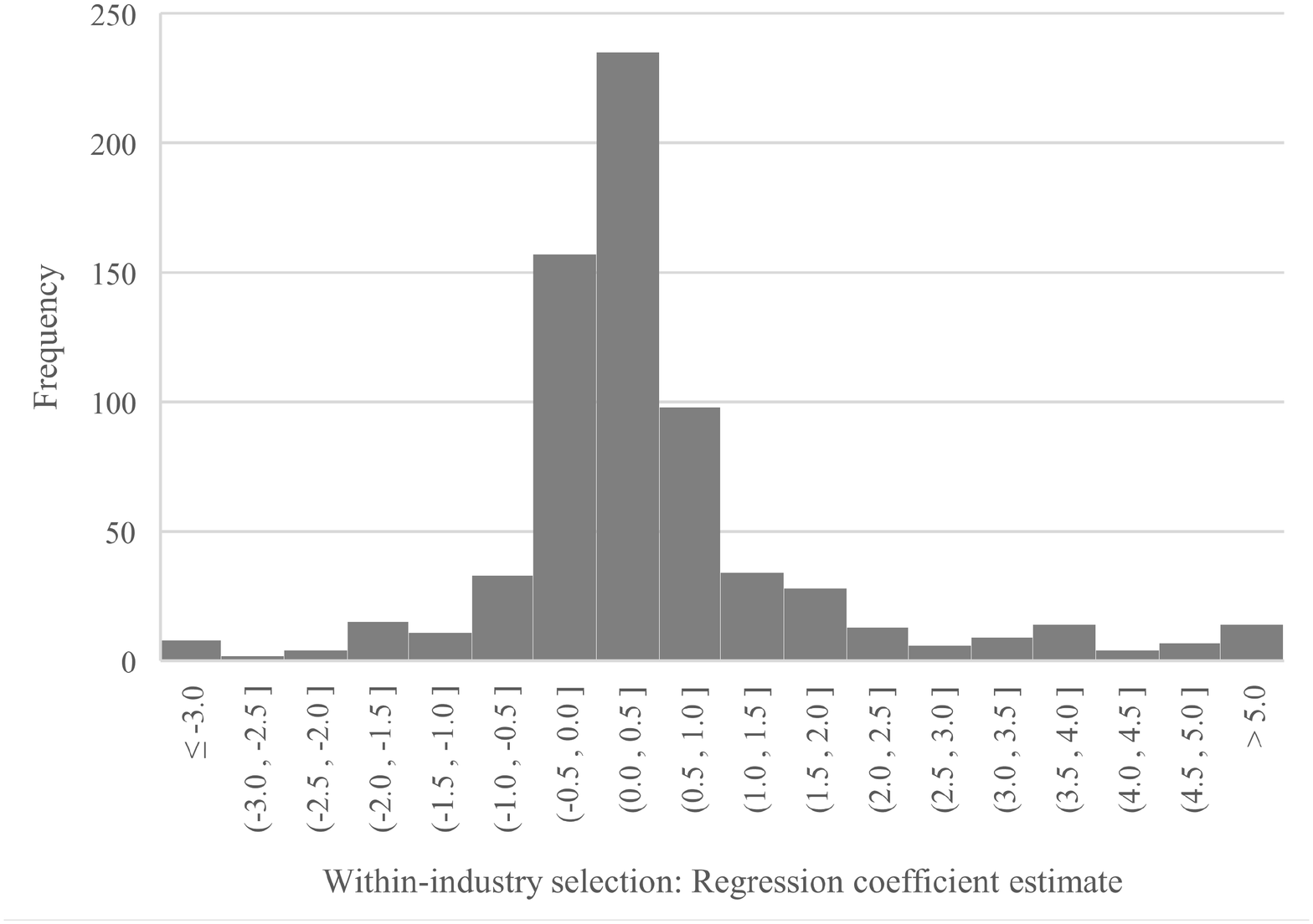}}
	\noindent\caption{}
	{\bf Histogram of fund's within-industry selection regression coefficient estimate under model (\ref{eq:persistence_check}).} 
	
	This figure describes the performance persistence of fund's within-industry selection. We consider stock funds with continuous performance data from 2015 to 2017. In this three-year sample, 692 stock funds are present.
	For each fund, we calculate the regression coefficient estimate in model (\ref{eq:persistence_check}) using data from 2015-2017 semi-annual reports. Then we plot histogram over all funds. \label{fig:15}
\end{figure}

\bigskip
\centerline{\bf [Place Figure~\ref{fig:15} about here]}
\bigskip

Second, in figure \ref{fig:15}, we examine the histogram of within-industry selection coefficient estimates for all funds. 
We can see that most funds show positive within-industry selection persistence. There are about 50 funds with regression coefficient estimate greater than 2. On the other hand, there are about 5 funds with regression coefficient estimate less than -2. For funds with positive within-industry selection persistence, we can use current within-industry selection to predict future within-industry selection. 

\begin{figure}[!htb]
	\centerline{\includegraphics[width=7in]{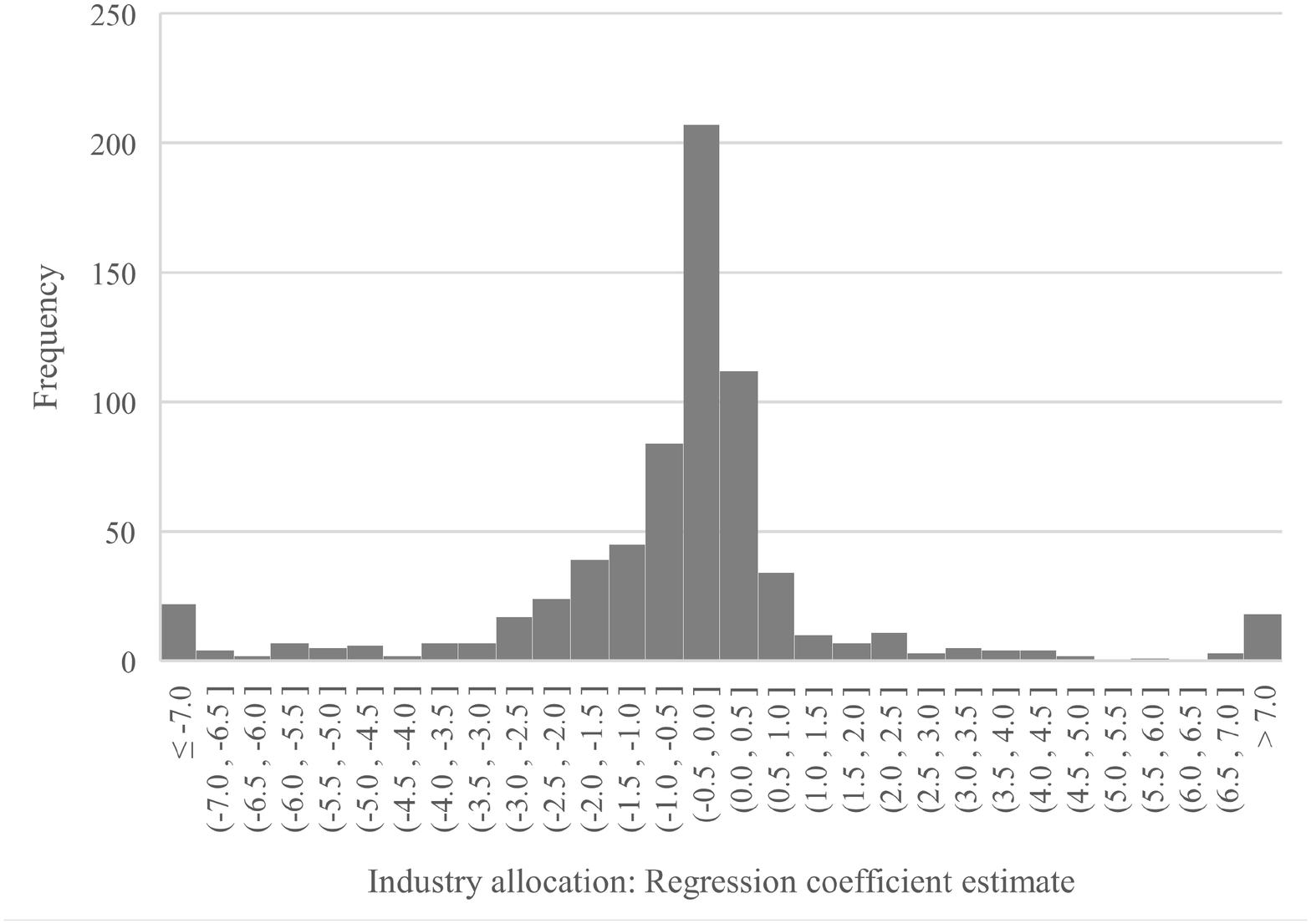}}
	\noindent\caption{}
	{\bf Histogram of fund's industry allocation regression coefficient estimate under model (\ref{eq:persistence_check}).} 
	
	This figure describes the performance persistence of fund's industry allocation. We consider stock funds with continuous performance data from 2015 to 2017. In this three-year sample, 692 stock funds are present.
	For each fund, we calculate the regression coefficient estimate in model (\ref{eq:persistence_check}) using data from 2015-2017 semi-annual reports. Then we plot histogram over all funds. \label{fig:16}
\end{figure}

\bigskip
\centerline{\bf [Place Figure~\ref{fig:16} about here]}
\bigskip

Third, for industry allocation, figure \ref{fig:16} presents the regression coefficient estimates for all stock funds. More funds show negative performance persistence than positive persistence. Tails of distribution are about symmetric. Most funds show coefficient estimate from -1.5 to 1. The performance persistence of industry allocation is not as good as within-industry allocation but better than asset allocation.

\begin{figure}[!htb]
	\centerline{\includegraphics[width=7in]{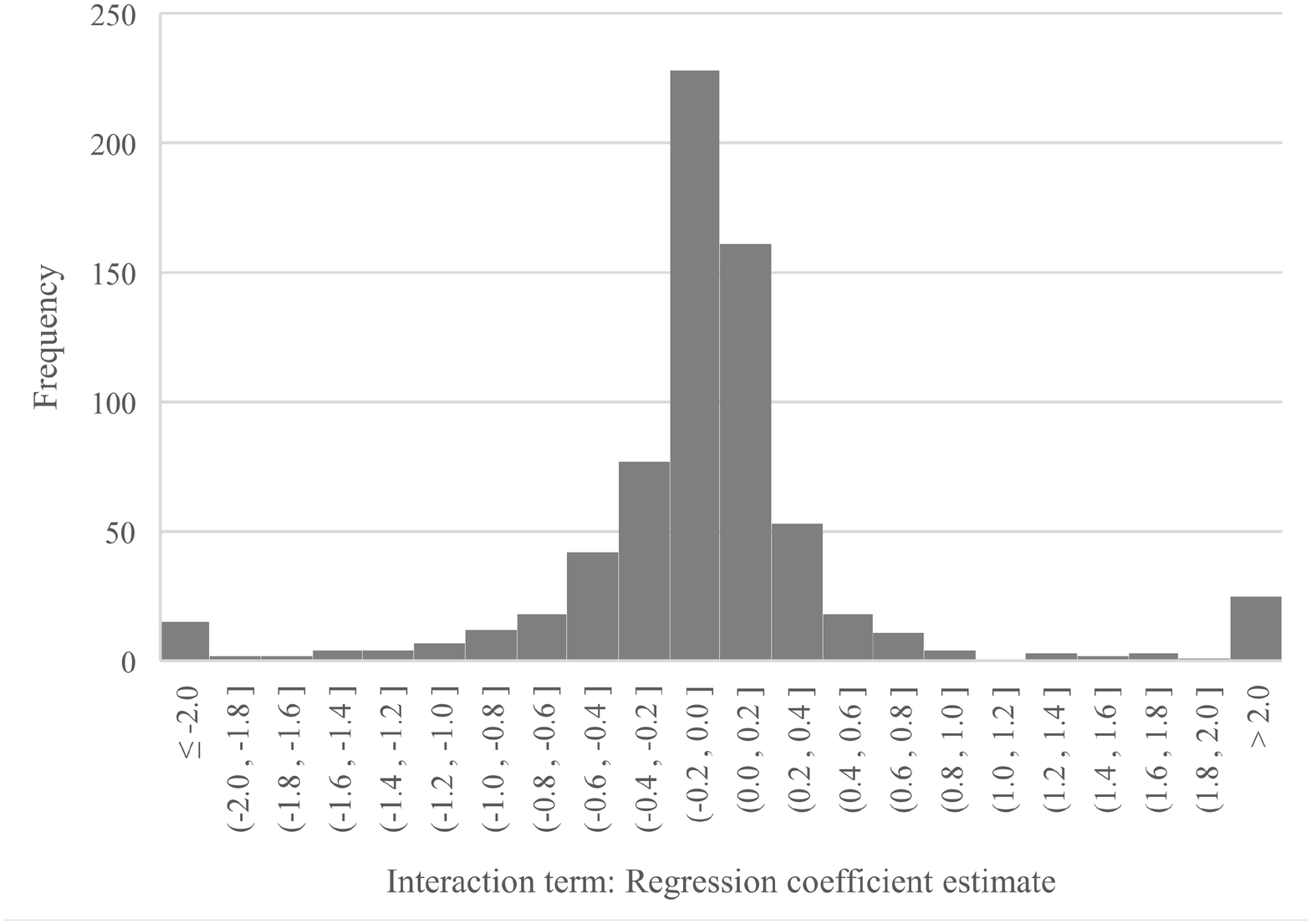}}
	\noindent\caption{}
	{\bf Histogram of fund's interaction term regression coefficient estimate under model (\ref{eq:persistence_check}).} 
	
	This figure describes the performance persistence of fund's interaction term. We consider stock funds with continuous performance data from 2015 to 2017. In this three-year sample, 692 stock funds are present.
	For each fund, we calculate the regression coefficient estimate in model (\ref{eq:persistence_check}) using data from 2015-2017 semi-annual reports. Then we plot histogram over all funds. \label{fig:18}
\end{figure}

\bigskip
\centerline{\bf [Place Figure~\ref{fig:18} about here]}
\bigskip

In previous section, we see that interaction term is mostly positive for all stock funds. Here we use model (\ref{eq:persistence_check}) to analyze the performance persistence of interaction term.
In figure \ref{fig:18}, we can see that regression coefficient estimates are mostly negative. That is, for more than 50\% of funds, their interaction term does not show positive persistence.

\bigskip
\centerline{\bf [Place Table~\ref{table:2} about here]}
\bigskip

Finally, we present summary in table \ref{table:2} about the performance persistence of within-industry selection, industry allocation, interaction term and asset allocation.
In model (\ref{eq:persistence_check}), we can test the hypothesis in (\ref{eq:persistence_check_hypothesis}) using t test. Rejection of null hypothesis indicates that performance persistence is significantly positive.
In our sample, we calculate the proportion of funds whose performance measure shows positive coefficient estimate, as in positive proportion from table \ref{table:2}.
We also calculate the proportion of funds whose performance measure shows significantly positive coefficient estimate, as in significantly positive proportion from table \ref{table:2}.

Among the performance measures we analyze, within-industry selection shows greatest positive proportion. There are 67\% of funds which show positive performance persistence of within-industry selection. Next to it is interaction term. There are 41\% of funds which show positive performance persistence of interaction term. 31\% of funds show positive performance persistence of industry allocation and 19\% show positive performance persistence of asset allocation.

As for significantly positive proportion, only 9\% of funds show significantly positive performance persistence in their within-industry selection. Only 1\% of funds show significantly positive performance persistence in their asset allocation. Among the performance measures, within-industry selection exhibits best performance persistence and asset allocation shows worst performance persistence.

\section{Association with Selection and Timing} \label{sec:Association}

In this section, we examine the association between within-industry selection and selection ability from Fama-French three-factor model. We also analyze the association between industry allocation and timing ability from Treynor-Mazuy model. We find the following results on the association between these measures from different models.

\bigskip
\centerline{\bf [Place Table~\ref{table:5} about here]}
\bigskip

In table \ref{table:5}, we present results found for the association between industry allocation and timing ability from Treynor-Mazuy model. We use six months before each semi-annual report to calculate return in the definition of industry allocation in (\ref{eq:ia}). We use monthly fund data to fit regression for Treynor-Mazuy model.
We consider funds with continuous performance in five years prior to the listed year. Sample size is the number of funds in the sample. Correlation is the Pearson correlation between industry allocation and timing ability. P value is for the Pearson correlation test. For positive correlation and p value less than 0.1, correlation between industry allocation and timing ability is significantly positive.

From table \ref{table:5}, we can see that in 2010-2014, 2011-2015 and 2012-2016 samples, the correlation between industry allocation and timing ability is significantly positive. 
However, in 2009-2013 sample, the correlation between industry allocation and timing ability is significantly negative. In 2008-2012 and 2013-2017 samples, there does not appear to be significant correlation between industry allocation and timing ability. These are samples when we do not see positive correlation between industry allocation and timing ability. To summarize, most of the time, there is positive correlation between industry allocation and timing ability.

\bigskip
\centerline{\bf [Place Table~\ref{table:6} about here]}
\bigskip

Then we study the association between within-industry selection and selection ability from Fama-French three-factor model. We use six months after each semi-annual report to calculate return in the definition of within-industry selection as in (\ref{eq:wiss}). We use monthly fund data to fit regression for Fama-French three-factor model. We consider funds with continuous performance in five years prior to the end year. 
For each fund, we calculate accumulated within-industry selection and fit regression model to get selection ability. 
Then based upon data for all funds, we calculate the Pearson correlation between within-industry selection and selection ability.

In table \ref{table:6}, we summarize the results of this association study. 
We can see that in all samples, within-industry selection and selection ability from Fama-French three-factor model are positively correlated. In all samples except 2010-2014, within-industry selection and selection ability show significantly positive correlation. In 2010-2014 sample, correlation is positive but not significant.

%

In this section, we conclude that selection ability from Fama-French three-factor model and within-industry selection defined in (\ref{eq:wiss}) are positively correlated, and that timing ability from Treynor-Mazuy model and industry allocation defined in (\ref{eq:ia}) are positively correlated most of the time.

\section{Conclusion} \label{sec:Conclusion}

In our paper, we use performance measures derived from Brinson model to analyze actively managed stock mutual funds in China.
We examine within-industry selection, industry allocation, interaction term and asset allocation. Within-industry selection depicts fund's ability to choose superior stocks within industry. Industry allocation describes fund's ability to choose superior industries. Interaction term shows whether funds over-weight both superior industries and superior stocks within industry, or over-weight both inferior industries and inferior stocks within industry. Asset allocation shows whether fund invests more in stocks in bulls and less in bears. 

Our conclusions are three-fold.
First of all, we conclude that most funds show positive within-industry selection but not industry allocation or asset allocation. Most funds show positive interaction term, which implies that their within-industry selection and industry allocation are positively associated. 
Second, among the performance measures we analyze, within-industry selection shows positive performance persistence in the greatest proportion of funds. Next to it is interaction term, and then industry allocation and asset allocation. 9\% of funds show significantly positive persistence in their within-industry selection. 
Finally, within-industry selection derived from Brinson model is significantly positively correlated with stock selection ability estimated from Fama-French three-factor model.
Industry allocation derived from Brinson model is positively correlated with timing ability estimated from Treynor-Mazuy model most of the time.

\bibliographystyle{unsrt}


\clearpage


\ 
\vfill
\begin{table}[!htb]
	\centerline{
		\begin{tabular}{@{}cccc@{}}
			\toprule
			Report Period & No. of  Funds & Average Size (10,000RMB) & Average No. of Stocks \\ \midrule
			Jun-03 & 2 & 89219 & 39 \\
			Dec-03 & 6 & 57577 & 70 \\
			Jun-04 & 7 & 82860 & 69 \\
			Dec-04 & 7 & 107254 & 46 \\
			Jun-05 & 9 & 85935 & 48 \\
			Dec-05 & 10 & 71664 & 40 \\
			Jun-06 & 11 & 47410 & 40 \\
			Dec-06 & 13 & 137276 & 41 \\
			Jun-07 & 15 & 392217 & 64 \\
			Dec-07 & 15 & 662911 & 75 \\
			Jun-08 & 15 & 354055 & 66 \\
			Dec-08 & 29 & 139758 & 43 \\
			Jun-09 & 40 & 210217 & 48 \\
			Dec-09 & 48 & 214452 & 60 \\
			Jun-10 & 53 & 152375 & 59 \\
			Dec-10 & 58 & 182026 & 63 \\
			Jun-11 & 62 & 157817 & 57 \\
			Dec-11 & 68 & 116231 & 50 \\
			Jun-12 & 75 & 106658 & 53 \\
			Dec-12 & 81 & 97773 & 49 \\
			Jun-13 & 90 & 81503 & 45 \\
			Dec-13 & 104 & 79870 & 47 \\
			Jun-14 & 143 & 67517 & 46 \\
			Dec-14 & 178 & 69810 & 43 \\
			Jun-15 & 264 & 96573 & 45 \\
			Dec-15 & 386 & 105265 & 45 \\
			Jun-16 & 541 & 65498 & 50 \\
			Dec-16 & 730 & 52228 & 63 \\
			Jun-17 & 990 & 47358 & 61 \\
			Dec-17 & 1087 & 41895 & 52 \\ \bottomrule
		\end{tabular}
	}
	\noindent\caption{}
	{\bf Summary of stock holdings data for actively managed stock funds in China.} 
	
	This table presents summary statistics of semi-annual stock holdings data for actively managed stock mutual funds in China. No. of funds is the number of stock mutual funds in the report period. Average size is the average stock holding capital of all stock funds at the report period. Average no. of stocks is the average number of stocks held by each fund at the report period. \label{table:1}
\end{table}
\vfill
\

\clearpage

\ 
\vfill
\begin{table}[!htb]
	\centerline{
		\begin{tabular}{@{}ccc@{}}
			\toprule
			& Model \ref{eq:benchmark_check} & Model \ref{eq:benchmark_reg1} \\ \midrule
			Greater than 1 & 0.59 & 0.73 \\
			Significantly not 1 at 10\% & 0.37 & 0.57 \\
			Significantly not 1 at 5\% & 0.28 & 0.50 \\ \bottomrule
		\end{tabular}
	}
	\noindent\caption{}
	{\bf Benchmark validation model in (\ref{eq:benchmark_check}) and (\ref{eq:benchmark_reg1}) for actively managed stock funds in China.} 
	
	This table presents summary statistics of benchmark validation model in (\ref{eq:benchmark_check}) and (\ref{eq:benchmark_reg1}) for actively managed stock mutual funds in China. We consider five-year sample covering all funds with continuous performance from 2013 to 2017. There are 487 funds present in five-year sample. Under model (\ref{eq:benchmark_check}) and (\ref{eq:benchmark_reg1}), we calculate regression coefficient estimate for each fund using data from 2013 to 2017. Greater than 1 is the proportion of funds whose regression coefficient estimate is greater than 1. Significantly not 1 at 10\% is the proportion of funds whose 90\% confidence interval of regression coefficient does not cover 1. Significantly not 1 at 5\% is the proportion of funds whose 95\% confidence interval of regression coefficient does not cover 1.
	\label{table:3}
	
\end{table}
\vfill
\

\clearpage

\ 
\vfill
\begin{table}[!htb]
	\centerline{
		\begin{tabular}{@{}ccc@{}}
			\toprule
			& Positive proportion & Significantly positive proportion \\ \midrule
			Industry allocation & 0.25 & 0.08 \\
			Within-industry selection & 0.72 & 0.40 \\
			Interaction term & 0.96 & 0.55 \\
			Asset allocation & 0.16 & 0.03 \\ \bottomrule
		\end{tabular}
	}
	\noindent\caption{}
	{\bf Proportion of funds whose average performance measure is positive.} 
	
	This table presents proportion of funds whose average performance measure is positive. We consider three-year sample covering all funds with continuous performance from 2015 to 2017. There are 692 funds in three-year sample. We calculate the average performance measure for each fund. Then we calculate the proportion of funds whose average performance measure is positive. We also use t test to see whether average performance measure of each fund is significantly positive or not. Significantly positive proportion is the proportion of funds whose average performance measure is positive at 10\% significance level. \label{table:4}
	
\end{table}
\vfill
\

\clearpage

\ 
\vfill
\begin{table}[!htb]
	\centerline{
		\begin{tabular}{@{}ccc@{}}
			\toprule
			& Positive proportion & Significantly positive proportion \\ \midrule
			Within-industry selection & 0.67 & 0.09 \\
			Industry allocation & 0.31 & 0.02 \\
			Asset allocation & 0.19 & 0.01 \\
			Interaction term & 0.41 & 0.03 \\ \bottomrule
		\end{tabular}
	}
	\noindent\caption{}
	
	{\bf Performance persistence for performance measures of actively managed stock funds in China.} 
	
	This table presents summary statistics of performance persistence for performance measures of actively managed stock mutual funds in China. We consider three-year sample covering all funds with continuous performance from 2015 to 2017.
	There are 692 funds present in three-year sample. Under model (\ref{eq:persistence_check}), we calculate regression coefficient estimate for each fund using data from 2015 to 2017. Positive proportion is the proportion of funds whose regression coefficient estimate is positive for the performance measure. Significantly positive proportion is the proportion of funds whose regression coefficient is significantly positive at 10\% significance level for the performance measure. \label{table:2}
	
\end{table}
\vfill
\

\clearpage

\ 
\vfill
\begin{table}[!htb]
	\centerline{
		\begin{tabular}{@{}cccl@{}}
			\toprule
			Year & Sample size & Correlation & P value \\ \midrule
			2012 & 191 & -0.0663 & 0.3619 \\
			2013 & 243 & -0.1809 & 0.0047*** \\
			2014 & 307 & 0.2158 & 0.0001*** \\
			2015 & 366 & 0.1809 & 0.0005*** \\
			2016 & 431 & 0.1153 & 0.0167** \\
			2017 & 487 & -0.0042 & 0.9257 \\ \bottomrule
		\end{tabular}
	}
	\noindent\caption{}
	{\bf Correlation between timing ability from Treynor-Mazuy mdoel and accumulated industry allocation from Brinson model.} 
	
	This table presents correlation between timing from Treynor-Mazuy model in (\ref{eq:Treynor_Mazuy}) and accumulated industry allocation from Brinson model defined in (\ref{eq:ia}). We consider rolling five-year sample covering all funds with continuous performance in five years. 
	In five-year sample, for each fund, we accumulate semi-annual industry allocation geometrically and use Treynor-Mazuy regression to calculate timing ability. 
	Then we compute the correlation between these two abilities using data from all funds in the sample.
	The column year is the end-year of the five-year sample. Sample size is the number of funds in the five-year sample. Correlation is the correlation between timing and accumulated industry allocation.
	P value shows how significant the correlation is.
	Here we use past six-month stock return to calculate performance measures at each semi-annual report.
	We use monthly data in five years to fit Treynor-Mazuy regression model.  \label{table:5}
	
\end{table}
\vfill
\

\ 
\vfill
\begin{table}[!htb]
	\centerline{
		\begin{tabular}{@{}cccl@{}}
			\toprule
			Year & Sample size & Correlation & P value \\ \midrule
			2012 & 191 & 0.3010 & 0.0000*** \\
			2013 & 243 & 0.4033 & 0.0000*** \\
			2014 & 307 & 0.0676 & 0.2377 \\
			2015 & 366 & 0.1505 & 0.0039*** \\
			2016 & 431 & 0.1821 & 0.0001*** \\
			2017 & 487 & 0.2769 & 0.0000*** \\ \bottomrule
		\end{tabular}
	}
	\noindent\caption{}
	{\bf Correlation between selection ability from Fama-French three-factor mdoel and accumulated within-industry selection from Brinson model.} 
	
	This table presents correlation between selection from Fama-French three-factor model in (\ref{eq:fama_french3}) and accumulated within-industry selection from Brinson model defined in (\ref{eq:wiss}). We consider rolling five-year sample covering all funds with continuous performance in five years. 
	In five-year sample, for each fund, we accumulate semi-annual within-industry selection geometrically and use Fama-French three-factor regression to calculate selection ability. 
	Then we compute the correlation between these two abilities using data from all funds in the sample.
	The column year is the end-year of the five-year sample. Sample size is the number of funds in the five-year sample. Correlation is the correlation between selection and accumulated within-industry selection.
	P value shows how significant the correlation is.
	Here we use future six-month stock return to calculate performance measures at each semi-annual report.
	We use monthly data in five years to fit Fama-French three-factor regression model.  
	\label{table:6}
	
\end{table}
\vfill
\



\ 
\vfill
\vfill
\


\ 
\vfill
\vfill
\

\clearpage

\ 
\vfill
\vfill
\

\ 
\vfill
\vfill
\

\ 
\vfill
\vfill
\

\clearpage

\ 
\vfill
\vfill
\

\ 
\vfill
\vfill
\

\ 
\vfill
\vfill
\

\clearpage

\ 
\vfill
\vfill
\

\ 
\vfill
\vfill
\

\ 
\vfill
\vfill
\

\ 
\vfill
\vfill
\

\ 
\vfill
\vfill
\

\ 
\vfill
\vfill
\

\ 
\vfill
\vfill
\

\ 
\vfill
\vfill
\

\ 
\vfill
\vfill
\

\ 
\vfill
\vfill
\

\clearpage

\ 
\vfill
\vfill
\

\ 
\vfill
\vfill
\

\ 
\vfill
\vfill
\

\ 
\vfill
\vfill
\

\end{document}